\documentclass[preprint,aps,nofootinbib]{revtex4}
\usepackage{graphicx}% Include figure files
\usepackage{dcolumn}% Align table columns on decimal point
\usepackage{bm}% bold math
\usepackage{epsf}
\usepackage{float}
\def\tlrho{\tilde\rho}

\def\Tc0{  { T_{c}^{0}    }  }
\def\INT01{  {   \dsint_{0}^{1} dq_1 dq_2 dq_3    }  }
\def\tiltc0{\tilde T_{c}^{0}}
\def\tilz{\tilde z}
\def\tile{\tilde E}
\def\tilp{\tilde P}
\def\tils{\tilde S}

\newcommand{\edc}{\end{document}}
\newcommand{\bb} {\begin{thebibliography}}
\newcommand{\eb}{\end{thebibliography}}
\newcommand{\bi}[1]{\bibitem{#1}}
\newcommand{\bc}{\begin{center}}
\newcommand{\ec}{\end{center}}

\newcommand{\be}{\begin{equation}\small}
\newcommand{\ee}{\end{equation}\normalsize}
\newcommand{\bea}{\begin{eqnarray}}
\newcommand{\eea}{\end{eqnarray}}
\newcommand{\ba}{\begin{array}{l}   }
\newcommand{\lab}[1]{\label{#1}}
\newcommand{\ea}{\end{array}}

\newcommand{\dsfrac}{\displaystyle\frac}
\newcommand{\ds} {\displaystyle}
\newcommand{\summa}{\ds\sum}
\newcommand{\dssum}{\summa}
\newcommand{\re}[1]{(\ref{#1})}

\newcommand{\tilT}{\tilde {T} }

\newcommand{\ci}{\cite}

\newcommand{\dsint}{\ds\int}

\def\bfr{{\bf r}}
\def\bfk{{\bf k}}

\newcommand{\epsq}{\varepsilon(q)}

\newcommand{{\vergul}}{  ,}

\newcommand{\veps}{\varepsilon }

\begin{document}
%\twocolumn
%\sloppy
\draft
%\doublespace
\title{Thermodynamics of noninteracting bosonic gases in cubic optical
lattices versus ideal homogeneous Bose gases}
\author{Abdulla Rakhimov$^{a,b}$}
%\email{rakhimovabd@yandex.ru}
\author{ Iman N. Askerzade $^{b,c}$}
\address{
$^a$ Institute of Applied Physics, National University of Uzbekistan, Tashkent 100174, Uzbekistan\\
$^b$ Computer Engineering  Department  of Ankara University, Ankara, 06100, Turkey\\
$^c$ Institute of Physics Azerbaijan National Academy of Sciences, AZ1143, Baku, Azerbaijan\\
}
%%%%%%%%%%%%%%%%%%%%%%%%%%%%%%%%%
\begin{abstract}

We have studied thermodynamic properties of noninteracting gases in periodic
lattice potential at arbitrary integer fillings and compared them with that of ideal homogeneous gases.
Deriving  explicit expressions for thermodynamic quantities and performing
exact numerical calculations we have found that the dependence of e.g.
entropy and energy on the temperature in the normal phase is rather weak
especially at large filling factors.
In the Bose condensed phase their power dependence on the reduced temperature is
nearly linear, which is in contrast to that of ideal homogeneous gases.
We evaluated the discontinuity in the slope of the specific
heat which turned out  to be approximately the same
as that of the ideal homogeneous Bose gas for filling factor $\nu=1$.
With increasing $\nu$ it decreases as the inverse of $\nu$.
  These results may  serve as a checkpoint
    for various experiments on optical lattices as well as theoretical studies of weakly interacting Bose
    systems in periodic potentials being a starting point for perturbative calculations.

\end{abstract}


\maketitle

%%%%%%%%%%%%%%%%%%%%%%%%%%%%%%%%%%%
\section{Introduction}
%%%%%%%%%%%%%%%%%%%%%%%%%%%%%%%%%%%%
Inter-particle interactions play a crucial role in fundamental physics. It is well known
that, coupling constants of interactions between elementary particles are set and
fixed by nature whereas that of atoms, especially in ultracold gases,
can be varied in a large scale by using Feshbach resonances \ci{feshbach}.
This gives an opportunity to change even the sign of the s- wave scattering length, $a_s$,
or to create a system of an ultracold Bose gas with extremely
weak interaction.
\footnote
{
As it is  known, there  is only one system of particles in nature, which approaches ideal gas regime at high densities, that  is a quark - gluon plasma of QCD.
}
 A good example is $^{39}K$. Recently \ci{roatiprl} the non-interacting
Bose - Einstein condensate (BEC) has been created  by sympathically
cooling a cloud of interacting $^{39}K$ atoms in an optical trap, and then $a_s$ is tuned
almost to zero by means of Feshbach resonance, in order to study effects of disorder.
In general, experimental and theoretical studies of ideal quantum gases can shed new light on the
interdisciplinary
phenomenon of Anderson localization \ci{Roatinature} and  a matter - wave interferometry \ci{8roati},
opening new directions towards   Heisenberg-limited  interferometry \ci{9roati}.
So, study of noninteracting Bose gases in optical traps \ci{bailier80}
or in optical lattices \ci{bailie80} has not only academic interest
 but also  a practical one.
In present work we study thermodynamics of noninteracting bosonic gases in cubic optical
lattices both in the BEC and normal phases.

Ultracold bosonic atoms in optical lattices have sparked investigations of strongly
 correlated many-body quantum phases with ultracold atoms
 \ci{lewenstein} that are now at the forefront of current researches.
 They may be used as quantum emulations of more complex condensed matter system,
and may greatly facilitate the achievements of the quantum Hall regime for the ultracold atomic gases
\ci{NCOOPER}.
Experimentally they are created  by superimposing two counter-propagating laser beams
of the same wavelength and frequency that act as an periodic potential. In the simplest
case, when the depth is constant and isotropic the potential can be represented as follows:
 \be
 V_{L}(\bfr)=V_{0}\sum_{\alpha=1}^{d}\sin^{2}(k_{0}^{\alpha}r_{\alpha}),\lab{1.1}
 \ee
where the wave vector $\bfk_{0}\equiv\{k_{0}^{\alpha}\}$ is related to the laser wavelength
$\lambda_{\alpha}$ as $k_{0}^{\alpha}=2\pi/\lambda_{\alpha}$, and $d$ is the space dimension
of a cubic lattice, $d=1,2,3$.

It is well known that, an ideal homogeneous  Bose (IHB) gas  of noninteracting atoms
consists of free atoms with the  plane wave
$\exp(i\bfk\bfr)$, and with the  energy dispersion relation $\epsilon_{k}=\bfk^{2}/2m$.
The creation of an optical lattice may be considered as a procedure of loading
preliminarily magnetically trapped ultracold Bose atoms into a well tuned
laser field, whose influence on the atoms, being in fact  the  Stark effect, is simulated via periodic potential \re{1.1}. Now the dispersion is no longer quadratic with
the momentum, but develops gaps at specific locations determined by the lattice structure.
This energy can be specified by a band index and a quasimomentum, taking on
values within the first Brillouin zone only.  As to the wave function it can be written as a Bloch
function
$\chi_{nk}(\bfr)=\sum_i \exp(i\bfk\bfr)\omega_n(\bfr-\bfr_i)$
 in the Wannier representation.
In the limit $V_0\gg E_R$, where $E_R$ is the recoil energy, each well of the periodic potential supports a number of vibrational
levels, separated by an energy $\omega_0\gg E_R$. At low temperatures, atoms are restricted to the vibrational level at each site. Their kinetic energy is then frozen, except
for the small tunneling amplitude to neighboring sites. The associated single particle
eigenstates in the lowest band are Bloch waves with quasimomentum
${\bfk}$ and energy
 \be
 \epsilon_0(\bfk)=3/2\omega_0-2J[\cos(k_x a)+\cos(k_y a)+\cos(k_z a)]+\ldots
 \lab{eps}
 \ee
where $\omega_0$ is the energy of local oscillations in the well \ci{zwerger}.
This is one of the main differences between IHB gas and noninteracting Bose gas
in optical lattices, which is no longer homogeneous either.
The bandwidth parameter $J>0$ is the gain in the  kinetic energy due to the nearest
neighbor tunneling, which can be approximated  as

\be
  J\approx\dsfrac{4}{\sqrt{\pi}}
  E_{R}\exp
  \left
  [-2\sqrt{\dsfrac{V_{0}}{E_{R}}}
     \right
    ] \left[
{\dsfrac{V_{0}}{E_{R}}}
\right ]^{3/4}
    \lab{2.1}
     \ee
where  $E_{R}= k_{0}^{2} / {2m}   $ , $k_0=\mid {\bf k}_0 \mid $ is the laser
 wave vector modulus, $k_0=\pi/a$, $a$ is the lattice spacing.

By the assumption that  the only lowest band is taken into account, an optical
lattice without harmonic trap can be described by the Bose-Hubbard model \ci{stoofbook},
\be
 \hat{H}=-J\sum_{\langle i,j\rangle}(\hat{b}_{i}^{+}\hat{b}_{j}+hc)+\frac{U}{2}\sum_{j}^{N_s}\hat{n}_{j}
  (\hat{n}_{j}-1),
  \lab{H}
   \ee
where $\hat{b}_{i}^{+}$   and $\hat{b}_{i}$ are the bosonic creation and annihilation operators on the site $ i$; the sum over $\langle i,j\rangle$ includes only
pairs of nearest neighbors; $J$ is the hopping amplitude, which is responsible for the tunneling of an atom from one
site to another neighboring site; $U$ is the on site repulsion energy, $\hat{n}_{j}$ is the number operator and $N_s$ the number of sites.
 Depending on the ratio $\kappa=U/J$ , the filling factor $\nu$ and the temperature $T$,  the system may be in superfluid,
 Mott insulator  or in normal $(N)$  phases. Note that, strictly speaking
 the Mott insulator  phase may be reached
 only for $T=0$ and  commensurate, i.e. integer filling factors \ci{Yukalovobsor}, $\nu=N/N_{s}$, where $N$ total
 number of atoms. The filling factor is related to
  the average atomic density, $\rho=N/V$ as $\nu=\rho a^d$, where $V$ - volume of the system occupied with the atoms.

  We shall study the system with $\kappa=0$
  described by the Hamiltonian  \re{H}  taking into account
only the lowest band with the dispersion \re{eps}.
   Since an ideal gas with the quadratic  spectrum at small momentum, i.e. $\epsilon(k)=ck^2+O(k^3)$  can not exhibit
   superfluidity \ci{landau}, our  discussions will concern  the phase transition from BEC
   phase into a normal phase.

On the other hand properties of $3d$ a  homogeneous  Bose  gas  of noninteracting atoms
are well known and  outlined in textbooks \ci{landau,huang}. Particularly, following
facts are well  established:
\begin{itemize}
\item
At sufficiently low temperatures they exhibit a phase transition from normal to Bose Einstein
state with the critical temperature
\be
  {\tilde T}_{c}^{0}=\dsfrac{2\pi}{m}\left[\dsfrac{\tlrho}{g_{3/2}(1)}  \right  ]^{2/3}
  \lab{tc23}
  \ee
  where $g_{\sigma}(z)=1/\Gamma(\sigma)\int_{0}^{\infty} x^{(\sigma-1)}
  dx/(\exp (x)z^{-1}-1)  $ is a Bose function, $m$ is the atomic mass. Here and below we denote
   by tilde the quantities characterizing the IHB gases and use the units $\hbar=1$, $k_B=1$. At the critical temperature such quantities as the
  energy - $E(T)$, entropy - S(T) and the specific heat, $C_v=(\partial E /\partial T)_v$
  are continuous, while the derivative of $C_v$ with respect to the temperature
  $\partial C_v/\partial T$ has a discontinuity given by
  \be
  {\tilde\Delta}= \dsfrac{{\tilde T}_{c}^{0}}{N}\left [ \left ( \dsfrac{\partial {\tilde C}_v}{\partial T}\right )_{   \tilde{T}_{c}^{-}} -
  \left (\dsfrac{\partial {\tilde C}_v}{\partial T}    \right)_{  \tilde{T}_{c}^{+}}\right]= 3.66
\lab{1.2}
  \ee
  where $N$ is the total number of particles.
\item
The critical temperature, ${\tilde T}_{c}^{0}$ , divides the scale of temperature into two different
regimes: BEC and normal. In the BEC phase, $T\leq {\tilde T}_{c}^{0} $
 when  the chemical potential of the gas is zero i.e. $\mu=0$,
 the energy and the specific heat
behave as ${\tilde E}/N \sim (T/{\tilde T}_{c}^{0})^{5/2}$, ${\tilde C}_v/N \sim (T/{\tilde T}_{c}^{0})^{3/2}$ which can be shown analytically \ci{huang}.
\item
The thermodynamic quantities exhibit a scale invariance, which means that they depend
explicitly on the reduced temperature $t=T/{\tilde T}_{c}^{0}$, but not on the density. One may find in the literature
good approximations for e.g. ${\tilde E}(t)$ or ${\tilde C}_v(t)$ \ci{wang}.
Moreover, there is a simple scale  relation between the internal  energy and the pressure
${\tilde P}$:
 \be
 {\tilde E}(t)=\dsfrac{d}{2} {\tilde P}(t)V
 \lab{pv}
 \ee
which holds for IHB gas exactly both in the BEC and normal phases \ci{mancarella}.
\end{itemize}

Now one may assume that  IHB gas is loaded into an optical lattice. Mathematically this means
that a periodic potential is implemented as an external potential given by Eq. \re{1.1}.
Clearly, the presence of this potential will change the energy dispersion and
makes a boundary for the quasimomentum $\bfk$. So, the bare dispersion is no longer quadratic
$\epsilon_k=\bfk^2/2m$ with $k=0 \div \infty$
 and the quasimomentum integration
 is taken only within the first Brillouin zone , $k_\alpha\in [-\frac{\pi}{a},\frac{\pi}{a}]$.  Then, how the properties of noninteracting Bose gas,
in particularly, listed above,
will be changed due to the periodic potential?  In present work we shall make an attempt to
answer these questions and make the parallel between the thermodynamic properties
IHB and optically confined  gases. The results will be useful in studying optical lattices
with a weak interaction and may serve as a check point for further theoretical studies.
\footnote{
Note that, there is one more system of Bose particles  whose
thermodynamic properties at finite temperature are similar to that of the gase in an optical lattice.
These are specific excitations in quantum antiferromagnets, namely, triplons, with a
non quadratic dispersion \ci{ourannals} for which the momentum integration is also taken in the Brillouin zone \ci{matsumoto} to study their possible Bose - Einstein condensation.}

We shall study high filling region $(\nu >3)$ also. This region is actual to investigate inteference and coherence properties of BEC, as well as properties of number squeezed states \ci{weli}.
Note that an experimental realization of optical lattice with high filling factor is rather difficult. Although
one dimensional optical lattice with $\nu=100$ \ci{polkovnikov} or array of condensates with 
$\nu=10^4$ has been created recently \ci{hajibaba} actual three dimensional opticall lattice
has no more than three atoms per site \ci{bloch}.

This paper is organized as follows. In Sec. II we derive main equations for thermodynamic
quantities of ideal optical lattices which will serve as a working formulas in the next sections. In Sec. III we discuss the power dependence of condensate fraction
on the reduced temperature. In Secs. IV and V the $T$ and $\nu$ dependence of the entropy
and energy will be studied. We study the heat capacity and its derivative in Sec.
VI. The Sec. VII will discuss the stability properties. Our conclusions are brought in
the last VIII section. In Appendix A and B we present useful  formulas for
IHB gases and ideal optical lattices respectively.

\section{Thermodynamic quantities: general relations }

The thermodynamic potential of the noninteracting Bose gas in an optical lattice is given by
\ci{ourknr2,ourknr1}
\be
\Omega=T\sum_q  \ln \left [  1-ze^{-\beta\epsq}    \right]
\lab{3.1}
\ee
where $\beta=1/T$, $\epsq=2J \dssum _{\alpha=1}^{3}(1-\cos\pi q_\alpha)$ - bare dispersion,
 $z=\exp(\beta\mu)$ - fugacity,  and
 $\sum_q f(q)\equiv N_s \int_{0}^{1}dq_1 dq_2 dq_3 f(q)$ .
 Note that in \re{3.1} the chemical potential $\mu$ includes the term $6J$, \ci{Yukalovobsor} which is not written here explicitly.

 From \re{3.1} one may find all needed thermodynamic quantities, namely
 \begin{itemize}
        \item {\bf Number of particles}:
        \be
        N=-\left ( \dsfrac{\partial \Omega}{\partial \mu }     \right)_T=
        \sum_q \dsfrac{1}{z^{-1}e^{\beta \epsq}-1       }
        \lab{4.1}
        \ee
        \item {\bf Entropy}:
        \be
        S=-\left ( \dsfrac{\partial \Omega}{\partial T }     \right)_\mu=
        -\beta
         \left[
        \Omega+\sum_q \dsfrac{(\mu-\epsq)}{z^{-1}e^{\beta \epsq}-1       }   \right ]
        \lab{4.2}
        \ee
        \item {\bf Energy:}
        \be
        E=TS+N\mu+\Omega= \sum_q \dsfrac{\epsq}{z^{-1}e^{\beta \epsq}-1       }
        \lab{4.3}
        \ee
  \item {\bf Pressure}:
  \be
  P=-\dsfrac{\Omega}{V}=-\dsfrac{\Omega}{a^d N_s}
  \lab{4.4}
  \ee
   \item {\bf Fugacity}.
         Actually in the BEC phase $z(T\leq T_{c}^{0})=1$. In the normal phase
        the function $z(T)$ may be found by solving \re{4.1} with  respect to $z$ for  given temperature T and filling factor.
        To calculate the specific heat one needs also $z'=(\partial z /\partial T)_{\rho}$.
        To obtain an explicit expression for $z'$ , we differentiate both sides of \re{4.1}
        for a fixed $N$ and solve the equation $d N/dT=0$ with respect
        to $z'$ \ci{pathria}. The  result is:
        \be
        z'=-\dsfrac{zR_{112}(z,T)}{TR_{012}(z,T)}
        \lab{4.5}
        \ee
          where we introduced following integrals
          \be
          \ba
          R_{ijk}(z,T)=\dsfrac{1}{N_s}\sum_q \dsfrac{
          (\beta \epsq)^i e^{j\beta \epsq}
          }
          {
          \left [ e^{\beta \epsq}z^{-1}-1 \right ]^k
          }
          \equiv \int_{0}^{1}dq_1 dq_2 dq_3
          \dsfrac{
          x^i e^{jx}
          }
          {
          \left [ e^{x}z^{-1}-1 \right ]^k
          },
          \\
          x=\beta \epsq
          \lab{5.1}
          \ea
          \ee
          Below we omit $T$ dependence of this function for simplicity, setting
          $R_{ijk}(z) \equiv R_{ijk}(z,T)$.
          Note that in the case of IHB gases the analogous integrals may be rewritten more
           compactly in terms of Bose functions $g_\sigma (z)$ due to the recurrent relations
            \re{A.3}.
    \item {\bf Heat capacity},
                  $C_v=(\partial E/ \partial T)_V$. Differentiating
 \re{4.3} and using \re{4.5}, \re{5.1} one obtains
      \be
     C_v/N_s=\left \{
     \begin{array}{ll}
		R_{212}(1) & (T\leq \Tc0) \\
\\
		 \dsfrac{R_{212}(z)R_{012}(z)-R_{112}^{2}(z)}{zR_{012}(z)}
            &  \quad (T>\Tc0)
	\end{array}
   \right.
          \lab{5.2}
  %  \ea
    \ee
    Note that, the heat capacity per particle $C_v/N$ may be evaluated by dividing
    these expressions to $\nu$ i.e. $C_v/N=[C_v/N_s]/\nu$.
         \item {{\bf The derivative of the specific heat}:  $(\partial C_v/\partial T)$}.
         From \re{5.2} one may obtain
         \be
         \dsfrac{T}{N_s}\left(\dsfrac{\partial C_v}{\partial T}\right)
         _{T\leq \Tc0} =2R_{323}(1)-2R_{212}(1)-
         R_{312}(1)
         \lab{5.3}
         \ee
        in the BEC phase and a rather long expression
        \be
        \ba
        \dsfrac{z^2T}{N_s}\left(\dsfrac{\partial C_v}{\partial T}\right)
         _{T > \Tc0}=2R_{323}(z)-2zR_{212}(z)-zR_{312}(z)+\\
         \\
         \dsfrac{R_{112}(z)[2zR_{112}(z)+3zR_{212}(z)-6R_{223}(z)]}{R_{012}(z)}
         -\\
         \\
         \dsfrac{2R_{112}^{2}(z)[zR_{112}(z)-3R_{123}(z)]}{R_{012}^{2}(z)}
         -2R_{112}^{3}(z)\left[\dsfrac{R_{023}(z)}{R_{012}^{3}(z)}\right]
         \lab{5.4}
        \ea
        \ee
        in the normal phase where we used Eqs. \re{4.5} and \re{5.1}.

             \end{itemize}
\section{The critical temperature and the condensed fraction}

\begin{figure}[h!]
%\begin{center}
\leavevmode
\includegraphics[width=0.7\textwidth]{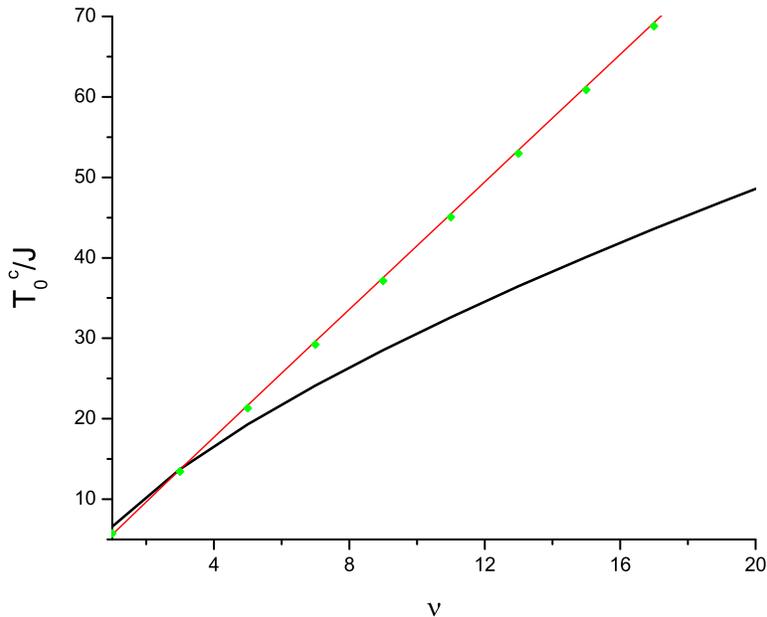}
%\end{center}
\caption{(Color online)
The condensation temperature in units $J$ as a function of the filling factor
calculated from Eq. \re{6.2} (thin solid line). The symbols correspond
to the exact solutions of Eq.\re{6.1}.
 The thick solid line corresponds to the critical temperature of
IHB  gas with an effective mass $m=1/2Ja^2$ and density $\rho=\nu/a^3$ exhibiting a power law $\nu^{2/3}$.
 }
\label{fig1tc0}
\end{figure}

For a noninteracting gas the phase transition BEC $\to N $ occurs when
$\mu\rightarrow 0$, so that the critical temperature $\Tc0$ is determined by the  following equation with a given filling factor $\nu_{c}$:
\be
        \nu_c=\int_{0}^{1}dq_1 dq_2 dq_3 \dsfrac{1}{e^{ \epsq/\Tc0}-1}
        \lab{6.1}
        \ee
which directly follows from Eq. \re{4.1}. The similar equation for IHB gas can be solved
analytically giving $\tilT_{c}^{0} \sim \rho^{2/3}$ power dependence as it was outlined above. However, for an ideal optical lattice the Eq.\re{6.1} can not
   be solved analytically, so  an explicit
dependence of $\Tc0(\nu)$ on the filling factor, which plays the role
 of density, $\nu=a^3\rho $ may be found
by studying numerical solutions of \re{6.1}.  We have recently shown
\ci{ouriman1} that the function $\Tc0(\nu)$ may be approximated as
\be
\Tc0(\nu)=3.96J\nu e^{0.37/\nu}
\lab{6.2}
\ee
   Particularly,    $\Tc0(\nu)$ has a linear dependence on  $\nu$ at large
   filling factors $(\nu \geq 5)$ as it is seen from Fig.\ref{fig1tc0}.
   % \footnote{
   %Here and in the most of next figures we plot the similar quantity for IHB gas,
%to make  a visual comparison between the both kind of gases.}
  This figure illustrates
   also the fact that the critical temperature  of an ideal gas is strongly modified
     by the  influence of the periodic potential, especially at large filling factors.
     Note that a magnetic trap with harmonic potential
     $U(r)=m(\omega_{x}^{2}x^2+\omega_{y}^{2}y^2+\omega_{z}^{2}z^2  )/2$
     also modifies the power dependence of critical temperature  as \ci{keterle}
     \be
     \Tc0(mag. trap)=0.94\omega_0 N^{1/3}
     \lab{traptc}
     \ee
where $\omega_0=(\omega_x \omega_y \omega_z)^{1/3}$
is the geometric average of trap frequencies.

\begin{figure}[h!]
%\begin{center}
\leavevmode
\includegraphics[width=0.7\textwidth]{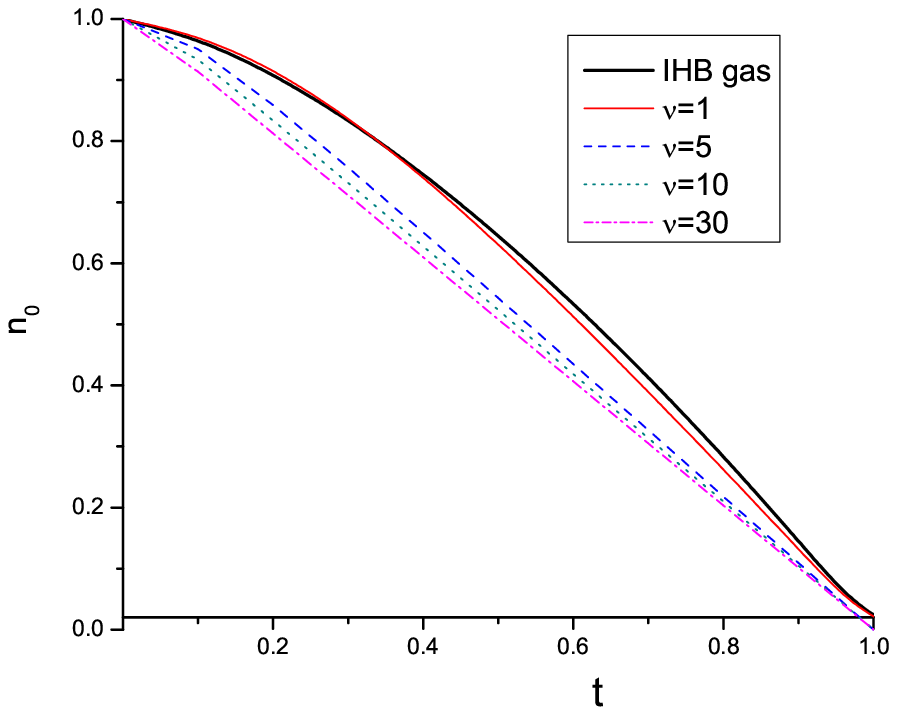}
%\end{center}
\caption{(Color online)
The condensed fraction $n_0$ vs reduced temperature $t=T/T_{c}^{0}$, as an exact solution of Eqs. \re{n0} and \re{6.3}
for  different values of $\nu$. The thick solid line corresponds to the
ideal homogeneous Bose gas given by \re{7.1}.
 }
\label{fig2n0}
\end{figure}

   The condensed fraction $n_0 =N_0/N$, where $N_0$ is the number of condensed particles,
    may be defined as
   \be
   n_0=1-n_1
   \lab{n0}
   \ee
   where
   \be
   n_1=\dsfrac{1}{\nu N_s}\sum_q \dsfrac{1}{e^{\beta \epsq}-1}=
   \dsfrac{1}{\nu}\INT01 \dsfrac{1}{e^{\beta \epsq}-1}
   \lab{6.3}
   \ee
%%%%%%%%%%%%% FIGURE 1 %%%%%%%%%%%%%%%%%%%%%%%%%%
 The function $n_0 (t,\nu)$, for various values of $\nu$ and the reduced temperature
 $t=T/\Tc0$  is presented in Fig.\ref{fig2n0}.
 %%%%%%%%%%%%% FIGURE 2 %%%%%%%%%%%%%%%%%%%%%%%%%%
It is seen from Fig.\ref{fig2n0} that at $\nu=1$ the temperature dependence the condensed fraction in both cases is almost the same:
\be
n_0(t,\nu)\mid _ {\nu=1} \sim {\tilde n}_0 (t)=1-t^{3/2}
\lab{7.1}
\ee
However, with the  increasing  of the filling factor, say, $(\nu \geq 5)$ , $n_0 (t,\nu)$ approaches to a
linear function of temperature i.e.
\be
n_0(t, {\nu \geq 5}) \sim 1-t.
\lab{7.11}
\ee

\section{Some scaling properties}
\subsection{Ideal homogenious Bose gas}
In general, in the thermodynamic limit,
$(N\rightarrow \infty$,  $ V\rightarrow \infty$,  $ \rho=N/V\neq\infty) $
 any thermodynamic quantity of IHB gas in the equilibrium is a function
of two independent variables, the temperature and the density , i.e. , $W=W(T,\rho)$ .

%{\bf Theorem.}
 It is easy to prove following statement:
 {\bf Let $W(T,\rho)$ of a nonrelativistic IHB gas has a dimensionality as $[W]=(Energy)^n (space)^{3l}$.
Then $W(T,\rho)$ may be represented as
\be
W(T,\rho)=(\tilde T_{c}^{0})^n\rho^{-l}{ W'}(t)
\lab{sc1}
\ee
where $W'(t)$ is the function of only reduced temperature, regardless of the particle mass $m$.
}\\
{\bf Proof }.
{\it
Since the most of quantities may be derived from the thermodynamic potential,
it is enough to prove this statement for $\Omega(T,\rho):$
\be
\Omega=VT\dsint \dsfrac{d{\bfk}}{(2\pi)^3} \ln \left [  1-ze^{-\veps(\bfk)/T}    \right]
\lab{sc2}
\ee
where $\veps(\bfk)=\bfk^2/2m$. Making  following  substitution in the integral
\be
k=\sqrt{2\tiltc0 mtx}, \quad \quad x=0 .. \infty
\lab{sc3}
\ee
 and integrating  by parts one obtains:
 \be
 \ba
 \Omega=-\dsfrac{4TN}{3\tlrho\lambda_{T}^{3}\sqrt{\pi}}\dsint_{0}^{\infty}
 \dsfrac{x^{3/2}dx}{e^x \tilz^{-1}-1}=-\dsfrac{NTg_{5/2}(\tilz)}{\tlrho \lambda_{T}^{3} }
  \lab{sc4}
 \ea
 \ee
where $\lambda_{T}=\sqrt{2\pi/mT}$ is the thermal wavelength,  and $\tilz$ is the fugacity.
On the other hand inverting Eq. \re{tc23} gives
\be
\tlrho\lambda_{T}^{3}=\zeta(3/2)t^{-3/2}
\lab{sc5}
\ee
where $\zeta (x)$ is the Riemann function.
Now inserting \re{sc5} into \re{sc4} we obtain
\be
\Omega=-\dsfrac{N\tiltc0 t^{5/2}g_{5/2}(\tilz)}{\zeta(3/2)}
\lab{sc6}
\ee
That is the dimensionless thermodynamic potential per particle $\Omega'/N$
may be presented \footnote{With $n=1$ and $l=0$ in Eq. \re{sc1}.} as a function only of $t$:
\be
\dsfrac{{ \Omega'}(t)}{N}=-\dsfrac{t^{5/2}g_{5/2}(\tilz)}{\zeta(3/2)}
\lab{sc7}
\ee
At the first
 glance it seems  that, the remaining explicit $\tlrho$ or $m$
dependence may come from $\tilz$. However, it can be shown that $\tilz$ satisfies
the above statement by itself: $\tilz(T,\rho)=\tilz(t)$. In fact, the equation
\be
\tlrho=\dsint \dsfrac{d{\bfk}}{(2\pi)^3} \dsfrac{1}{\tilz^{-1}e^{\veps(\bfk)/T}-1}
\lab{sc8}
\ee
which defines   $\tilz$, may be, clearly, rewritten as
\be
\tlrho\lambda_{T}^{3}=g_{3/2}(\tilz)
\lab{sc9}
\ee
Now from Eqs. \re{sc5} and \re{sc9} we get
\be
g_{3/2}(\tilz)=\zeta(3/2)t^{-3/2}
\lab{sc10}
\ee
Thus $\tilz(T,\rho)=\tilz(t)$, and hence the density dependence of  e.g. $\Omega'$
is involved
only via the reduced temperature as $t\sim T/\tlrho^{2/3}$.
}
%%%%%%%%%%%%% FIGURE 3 %%%%%%%%%%%%%%%%%%%%%%%%%%
\begin{figure}[h!]
%\begin{center}
\leavevmode
\includegraphics[width=0.7\textwidth]{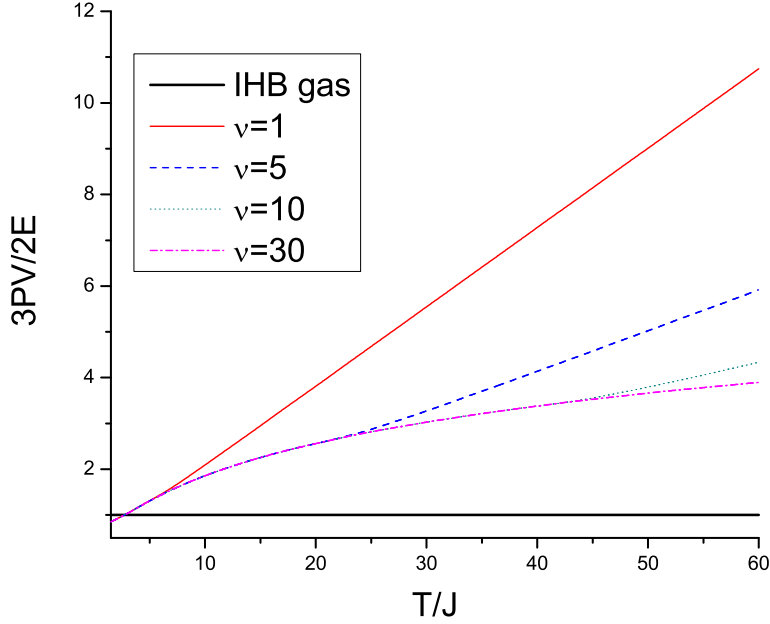}
%\end{center}
\caption{(Color online)
This relation between the internal energy and the thermodynamic potential,
$\mid\Omega\mid$, equals exactly to unity for ideal homogeneous Bose gas, as presented by  the thick solid line (see Eq.s \re{omegagaz} and \re{A.8} )
 }
\label{fig3pv2e}
\end{figure}

%%%%%%%%%%%%% FIGURE PVTNS %%%%%%%%%%%%%%%%%%%%%%%%%%
%\begin{figure}[h!]
%\begin{center}
%\leavevmode
%\includegraphics[width=0.7\textwidth]{PVTNS.eps}
%\end{center}
%\caption{(Color online)
%Quantity $PV/TN_s$
% }
%\label{figpvtns}
%\end{figure}

Note that in the above discussions we didn't lose the number of free parameters. Actually,
there are two independent variables: on the LHS of \re{sc1} they are $T$ and $\tlrho$, while
on the RHS $T$ and $\tiltc0$.

Similarly, it can be easily shown that
\be
\ba
\dsfrac{\tile ' (t)}{N}=\dsfrac{3 t^{5/2}g_{5/2}(\tilz) }{2\zeta(3/2)}
\\
\\
\dsfrac{\tils ' (t)}{N}=\dsfrac{5t^{3/2}g_{5/2}(\tilz) }{2\zeta(3/2)}
\\
\\
{\tilp '  (t)}=\dsfrac{t^{5/2}g_{5/2}(\tilz) }{\zeta(3/2)}
\lab{sc11}
\ea
\ee
where $\tilp (t)'= {\tilp}/\tiltc0 \rho$ (see Appendix A).
\subsection{Ideal optical lattice}
The natural question arises, if a thermodynamic quantity of an ideal optical lattice
satisfies the scale relation given in \re{sc1}, where the role of the density plays
the filling factor $\nu=\rho a^3$? To answer this question we consider the BEC and
the normal phases separately.

In the BEC phase the thermodynamic potential per site is given by
\be
\Omega(T\leq \Tc0)/N_s=T a^3\dsint_{-\pi/a}^{\pi/a}\dsfrac{dk_1dk_2dk_3}{(2\pi)^3}
\ln \left [  1-ze^{-\veps(\bfk)/T}    \right] _{z=1}
\lab{sc12}
\ee
where $\veps(\bfk)=2J\dssum_{\alpha=1}^{3} (1-\cos(k_\alpha a))$.
The Eq. \re{sc12} displays that in this phase  $\Omega(T)/N_s$ absolutely does not dependent on $\nu$.
So, it can be represented as    a function of the reduced temperature $t$ as:
 $\Omega(T\leq \Tc0)/N_s= \Omega(t\Tc0)/N_s$ and hence the thermodynamic potential per particle behaves as $\Omega(T)/N\sim 1/\nu$, due to the relation $\nu N_s=N$.

In the normal phase an explicit $\nu $ dependence of all thermodynamic quantities
comes from the function $z(t,\nu)$. To illustrate the fact  that in contrast to the fugacity of
IHB gas,  the fugacity  of a noninteracting  gas in the periodic potential
  dependence not only on
$t$ but also on $\nu$ we will use a simple Debay approximation (see Appendix B).
So, on the one hand
\be
\nu=a^3\dsint_{-\pi/a}^{\pi/a}\dsfrac{dk_1dk_2dk_3}{(2\pi)^3}\dsfrac{1}{z^{-1}e^{\veps(k)/T}-1}
\simeq a^3 \dsint_{0}^{k_D}\dsfrac{k^2dk}{2\pi^2}\dsfrac{1}{z^{-1}e^{\veps_k/T}-1}
\lab{sc13}
\ee
 where $\veps_k=\bfk^2/2m$, $m=1/2Ja^2$, $k_D=(6\pi^2)^{1/3}/a$.
 On the other hand $\Tc0$
corresponding to  this fixed $\nu$ is defined by
 \be
\nu\simeq a^3 \dsint_{0}^{k_D}\dsfrac{k^2dk}{2\pi^2}\dsfrac{1}{e^{\veps_k/\Tc0(\nu)}-1}
\lab{sc14}
\ee
Now equating the last  two equations to each other and making following substitution
\be
k=\sqrt{2mx\Tc0} ,\quad \quad x=0..x_D, \quad \quad x_D=\dsfrac{6^{2/3}\pi^{4/3} }
{2a^2m\Tc0(\nu)}
\lab{sc15}
\ee
one obtains following equation with respect to $z$
\be
\dsint_{0}^{x_D}\sqrt{x}dx\left[
\dsfrac{1}{z^{-1}e^{x/t}-1}-\dsfrac{1}{e^{x}-1}
\right ]=0
\lab{sc16}
\ee
Although the integrand in Eq. \re{sc16} depends only on $t$
the upper boundary of the integral explicitly depends on $\nu$ e.g. through Eq. \re{6.2}
as $\Tc0\approx4\nu J$ and as a result $z$ acquires an explicit, nonlinear  $\nu$ dependence.

Thus we may conclude that a thermodynamic quantity describing an ideal optical lattice
may be presented as the only function of $T$  in the BEC phase, while in the normal phase,
its density dependence cannot be simply extracted as it was done for IHB gas in Eq. \re{sc1}.

In the content of the scaling, there is a simple scale  relation between the internal  energy and the pressure
${\tilde P}$:
 \be
 {\tilde E}(t)=\dsfrac{d}{2} {\tilde P}(t)V
 \lab{pv}
 \ee
which holds for IHB gas exactly. In textbooks it is usually derived by the integration by parts
 of the free energy, ${\tilde \Omega}=-{\tilde P}V$. On the other hand, it can be
  shown that  \ci{mancarella}, this relation is a consequence
of the scale invariance of the Hamiltonian with respect to the dilation of coordinates such as
$r\rightarrow \lambda r$.

To discuss the scaling relation $E=3PV/2$ we plot in Fig.\ref{fig3pv2e} dimensionless quantity
 $3PV/2E$ which equals exactly  to unity for IHB gases. It is seen from Fig.\ref{fig3pv2e}
 that the presence of the external potential leads to a strong  breaking of this relation irrespective
  of the values of  $\nu$  and $t$.
%%%%%%%%%%%%% FIGURE 4  A,B,C,D %%%%%%%%%%%%%%%%%%%%%%%%%%
\begin{figure}[H]
\begin{minipage}[h]{0.49\linewidth}
\center{\includegraphics[width=1.1\linewidth]{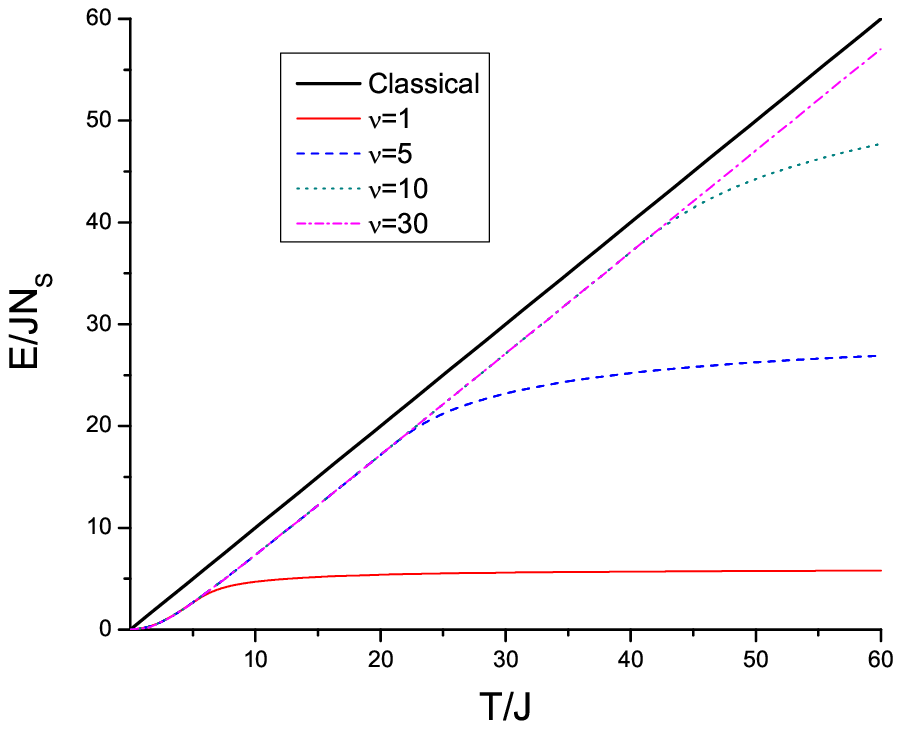} \\ a)}
\end{minipage}
\hfill
\begin{minipage}[h]{0.49\linewidth}
\center{\includegraphics[width=1.1\linewidth]{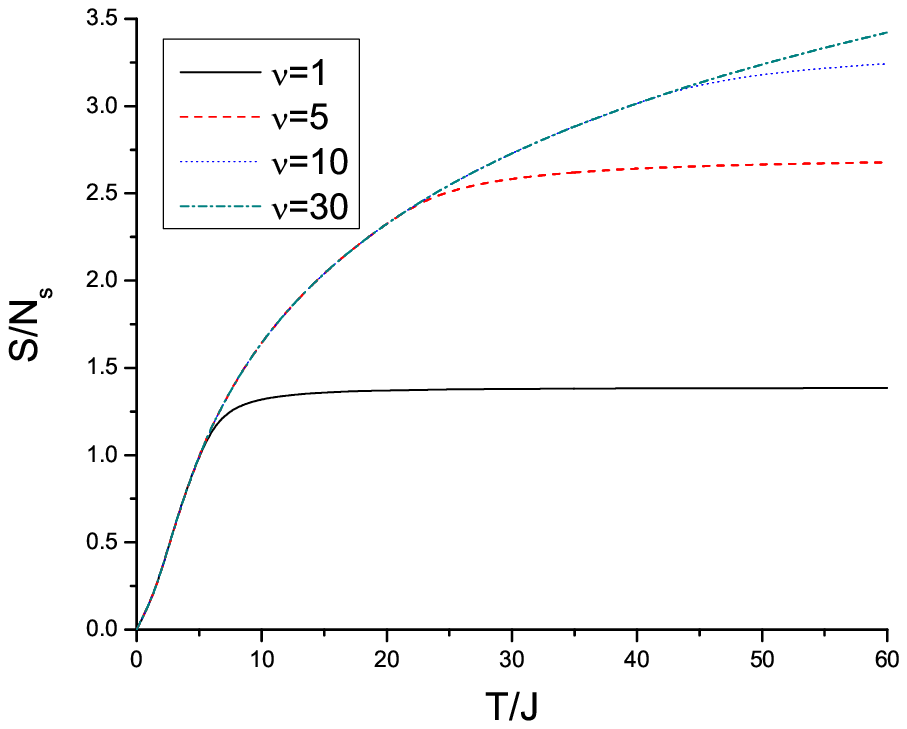} \\ b)}
\end{minipage}
\medskip
\medskip
\medskip
\begin{minipage}[H]{0.49\linewidth}
\center{\includegraphics[width=1.1\linewidth]{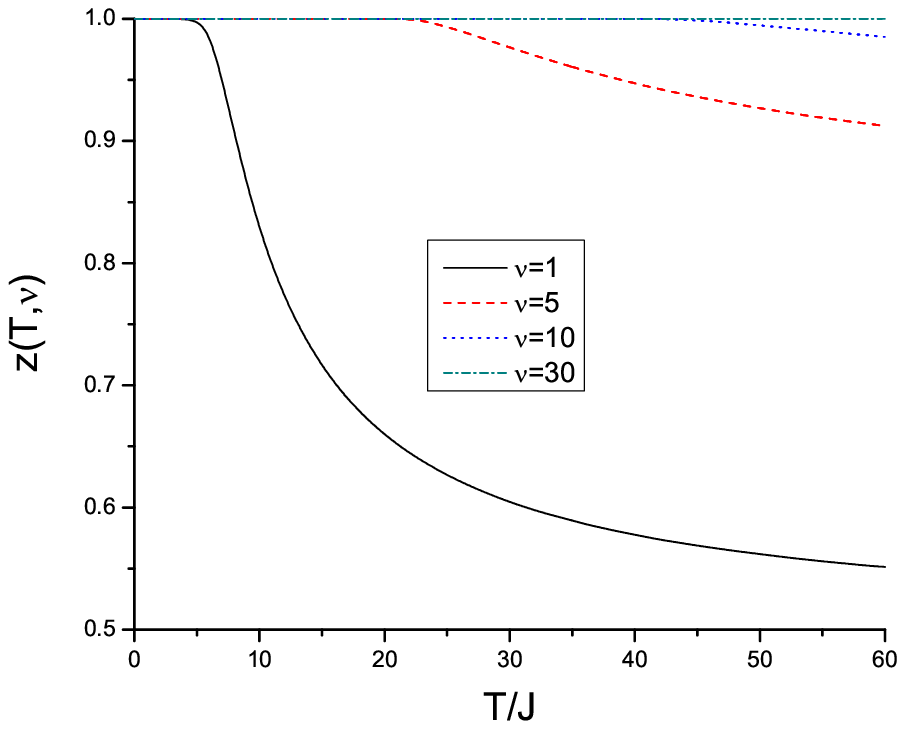} \\ c)}
\end{minipage}
\hfill
\begin{minipage}[H]{0.49\linewidth}
\center{\includegraphics[width=1.1\linewidth]{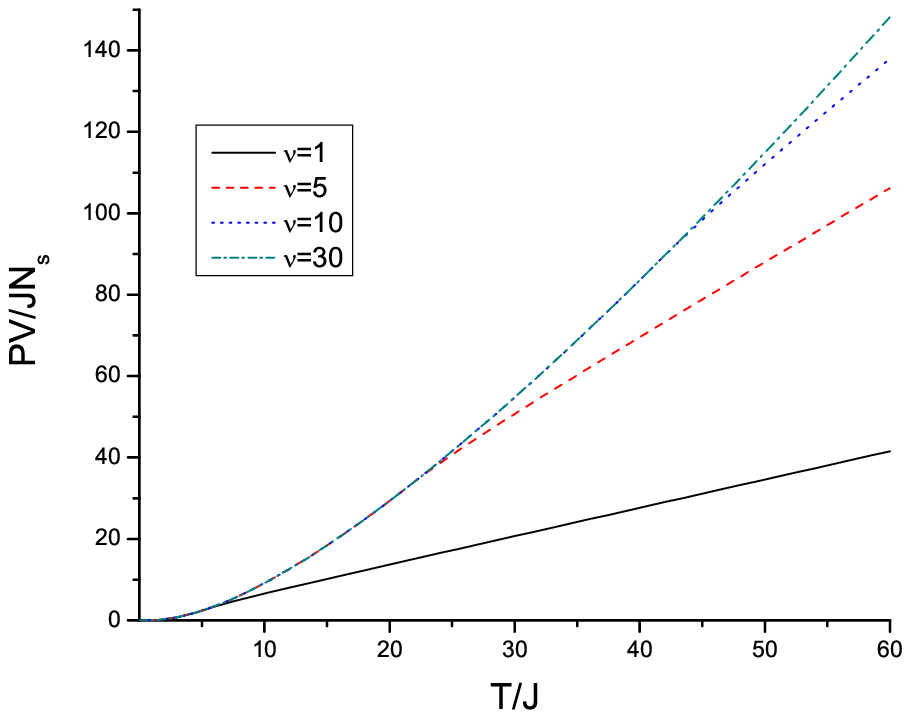} \\ d)}
\end{minipage}
\caption{(Color online)
The energy per site in units $J$, (a), the entropy per site (b), the fugacity   (c)
and the absolute value of the thermodynamic potential per site in units $J$, (d) of noninteracting gases in optical gases
for different  filling factors $\nu$
vs the absolute scale of the temperature $T/J$. The branch points correspond to the
following critical temperatures: $\Tc0(1)=5.5911J$, $\Tc0(5)=21.6714J$  and
 $\Tc0(10)=41.5J$  .
  \label{fig4abcd}}
\end{figure}
%%%%%%%%%%%%%%%%%%%%%%%%%%%%%%%%%%%%%%%%%%%%%%%%%%%%%%%

\section{$T$ and $\nu$ dependence of thermodynamic quantities }

To discuss the general tendency of a thermodynamic quantity of an ideal optical lattice
as a function of temperature and filling factor we present in Figs. \ref{fig4abcd}
the energy (\ref{fig4abcd}a), entropy (\ref{fig4abcd}b), fugacity
(\ref{fig4abcd}c) and the absolute value of thermodynamic potential per site
  in units $J$ (\ref{fig4abcd}d) on the scale of the dimensionless absolute temperature for various $\nu$.
The  branch points correspond to the
following critical temperatures: $\Tc0(1)=5.5911J$, $\Tc0(5)=21.6714J$  and
 $\Tc0(10)=41.5J$  which separate the BEC and the normal phases.
   Below we consider these two phases separately.

 a){\it The BEC phase.}
  In this phase these quantities does not depend on $\nu$
 as it was shown in the previous section. It is seen from Figs. \ref{fig4abcd}
 that for $T\leq \Tc0$ the power dependence of $E(T)$, $S(T)$ and $P(T)$ on temperature is nearly linear  e.g. $E(T)\sim T$ . This is similar to the classical gas, as it is illustrated
  in Fig. \ref{fig4abcd}(a) and in contrast to the case of IHB gas where
 for instance $S/N\sim T^{3/2}$ (see Appendix A).

b){\it The normal phase.}
The difference between the thermodynamic quantities of noninteracting particles in periodic potential and that of IHB gas is rather large for $T>\Tc0$. As it is seen from
Figs.~\ref{fig4abcd} (a) and (b) when the temperature reaches its critical value,
$\Tc0(\nu)$, the energy and the entropy become quite insensitive to temperature,
especially at large filling factors $\nu\geq 5$.
 Mathematically this tendency  may be explained as follows. With the increasing of temperature the exponential function in Eq. \re{4.3} fast decreases. On the other
hand as it is clearly seen from Fig.~\ref{fig4abcd}(c) the factor $z^{-1}$ in this equation
also increases exponentially (see Footnote 3). As a result the whole product
$\exp(\veps(q)/T)/z(T)$ in Eq. \re{4.3} goes to a constant value at large $T$
and $\nu$
as it is displayed in Fig.5(d). For the similar reason
the absolute value of the dimensionless thermodynamic potential per site, $|\Omega| /JN_s$ plotted in Fig.\ref{fig4abcd}(d) increases with the increasing of the temperature, as
it is clear from the Eq. \re{3.1}. Note also that, the pressure (Fig. 4(d))  and module of the chemical
potential $|\mu |$ fast increases (Fig. 5(c)) with the increasing of temperature and the density as expected from general  physical principles.
%However, the above found  fact that
%the whole internal energy of the system in the normal phase
% remains almost unchanged   during the heating, seems rather  unphysical.
 %Although such a prediction  can be, in principal,
  %verified experimentally in an optical lattice in $^{39}K$. On the other hand
   %this failes it is may be  caused
  %by our simple approximation, which neglects  higher bands,   is unacceptable
  %for high filling factors and temperature.

\begin{figure}[H]
\begin{minipage}[H]{0.49\linewidth}
\center{\includegraphics[width=1.1\linewidth]{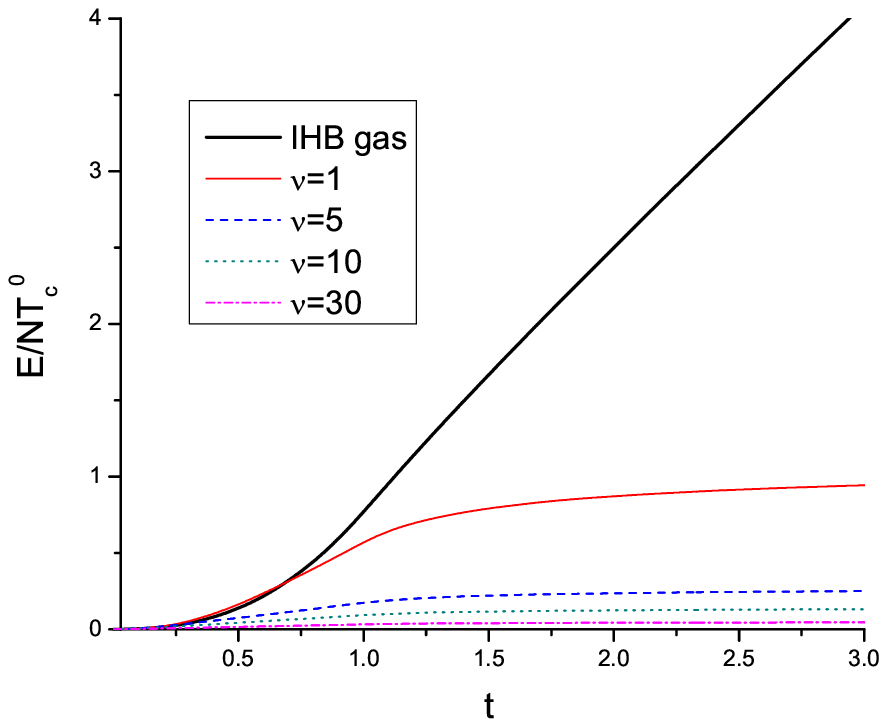} \\ a)}
\end{minipage}
\hfill
\begin{minipage}[H]{0.49\linewidth}
\center{\includegraphics[width=1.1\linewidth]{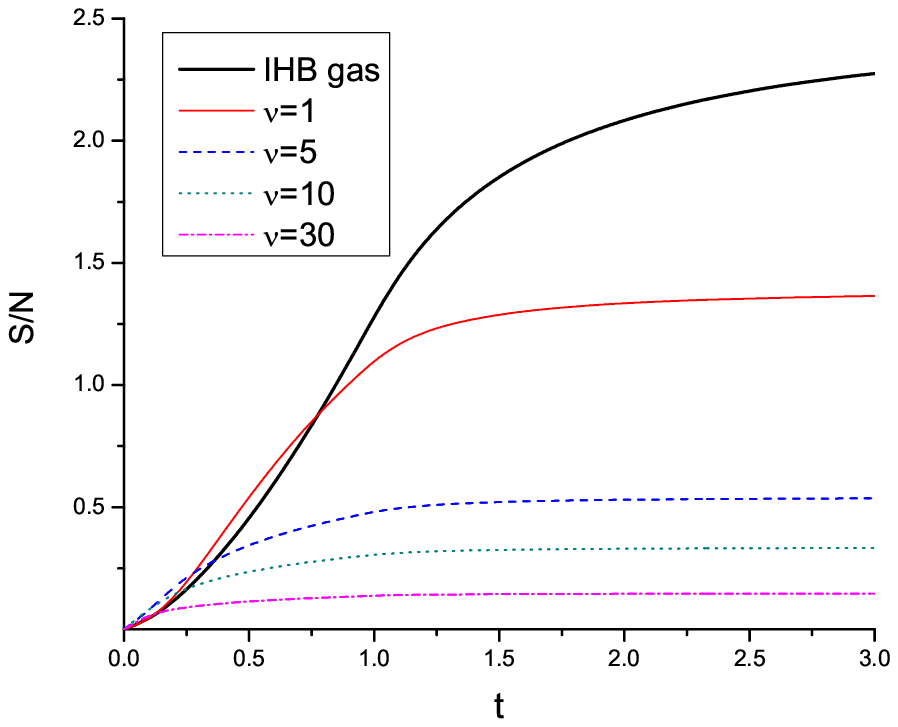} \\ b)}
\end{minipage}
\medskip
\medskip
\medskip
\begin{minipage}[H]{0.49\linewidth}
\center{\includegraphics[width=1.1\linewidth]{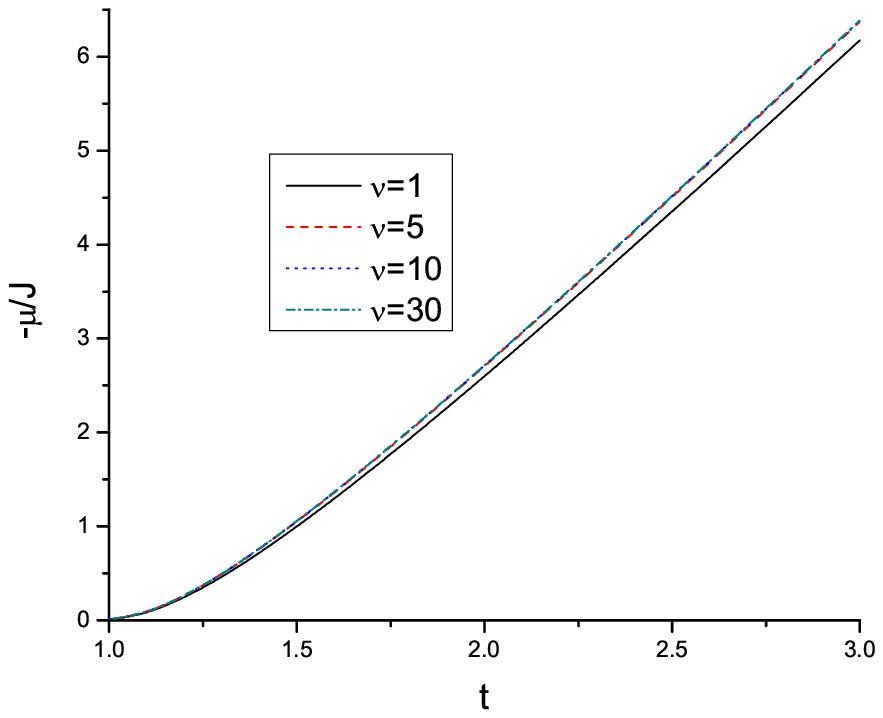} \\ c)}
\end{minipage}
\hfill
\begin{minipage}[H]{0.49\linewidth}
\center{\includegraphics[width=1.1\linewidth]{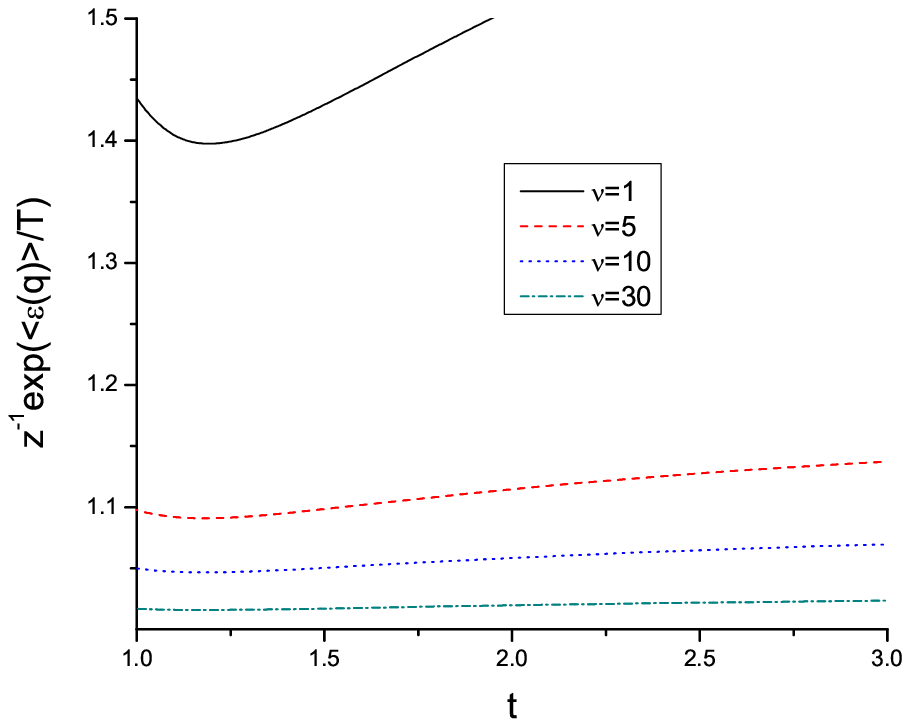} \\ d)}
\end{minipage}
\caption{(Color online)
 The energy in units $\Tc0$  (a),  entropy (b) per particle and the absolute value
 of the chemical potential in units $J$ (c)
for  different values of $\nu$. In the evaluating of the quantity
 $z^{-1}\exp(<\veps(q)/T>)$ presented in (d) we set $<\veps(q)=1>$. The thick solid lines in (a) and (b) correspond to the
ideal homogeneous Bose gas given by \re{A.7} and \re{A.8}.
\label{fig5ab}}.
\end{figure}

To give a further illustration of  the contrast between IHB gas and the one in the periodic potential
in the normal phase we present in Figs. \ref{fig5ab} the energies  (a) as well as entropies
(b) of both systems on the scale of the reduced temperature. As it is seen from the figures,
with the increasing  the temperature , for instance, the energy of the IHB gas continues  to increase  , while that of the ideal optical lattice remains nearly constant.
For small filling factors $S(t,\nu)/N$ is almost
 linear in the temperature. As for the $\nu$ dependence of this function, the entropy
 decreases as $1/\nu$ with increasing $\nu$. A similar behavior of $S(t,\nu)/N$
 was found in Ref. \ci{pra69} where the authors studied entropy -
 temperature curves in a large scale of temperature for $\nu \leq 4$.

 It is clear that interatoimic interaction  become crucial at large
filling factors. Thus our results for  $\nu > 4$ may be considered just as
model calculations.

%%%%%%%%%%%%% FIGURE 5A,B %%%%%%%%%%%%%%%%%%%%%%%%%%
%%%%%%%%%%%%%%%%%%%%%%%%%%%%%%%%%%%%%%%%%%%%%%%%%%%%%%
%%%%%%%%%%%%%%%%%%%%%%%%%%%%%%%%%%%%%%%%%%%%%%%%%%%%%%%

 \section{The specific heat and the jump in $\partial C_v/\partial T$}

\begin{figure}[H]
\begin{minipage}[H]{0.49\linewidth}
\center{\includegraphics[width=1.1\linewidth]{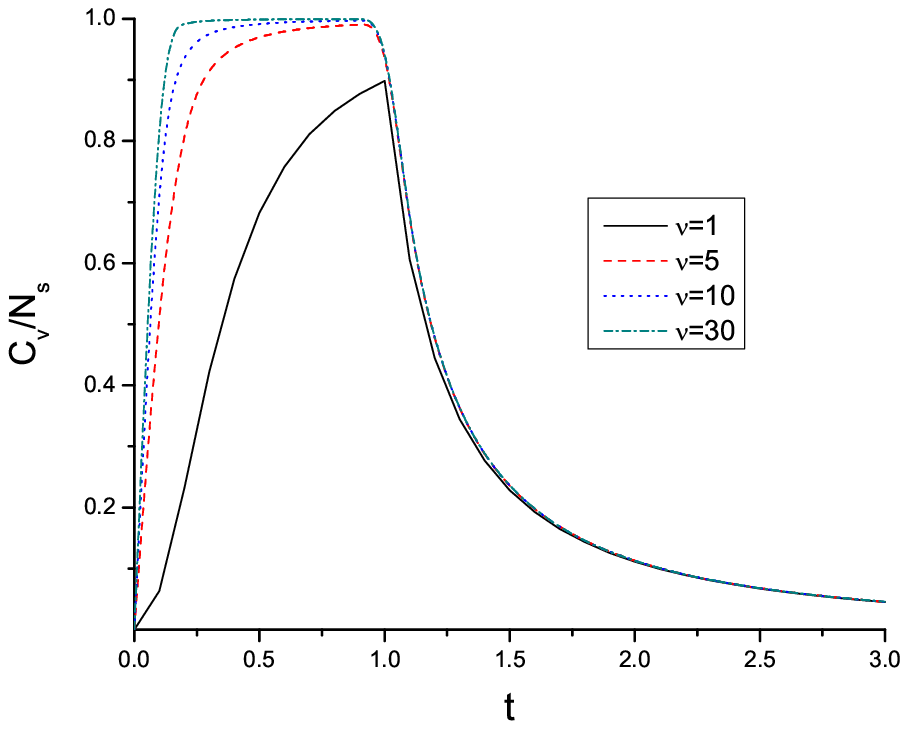} \\ a)}
\end{minipage}
\hfill
\begin{minipage}[H]{0.49\linewidth}
\center{\includegraphics[width=1.1\linewidth]{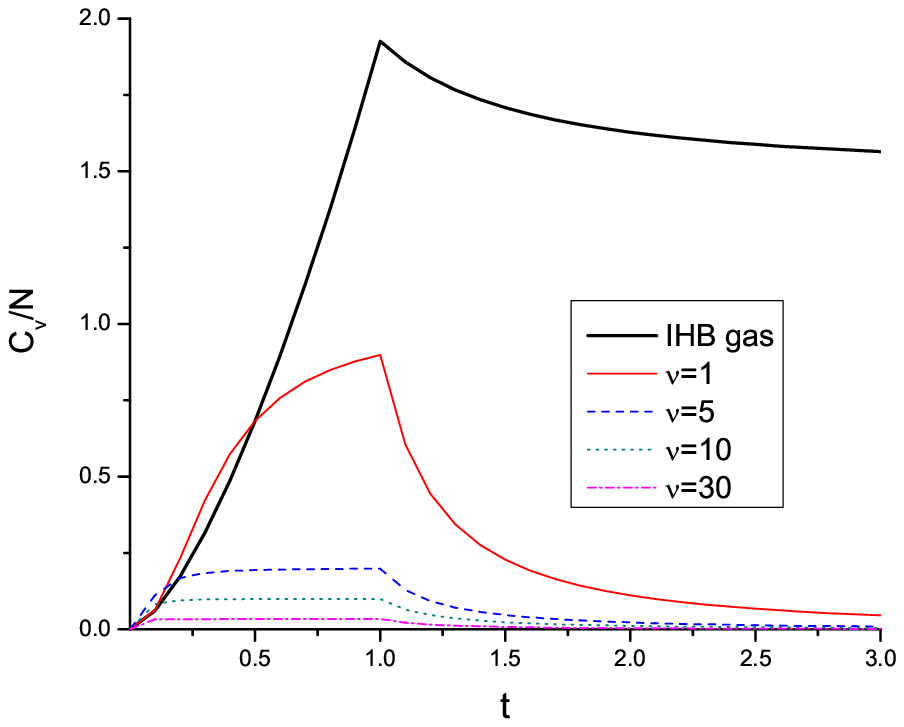} \\ b)}
\end{minipage}
\caption{(Color online) The specific heat per site (a), evaluated from
Eq. \re{5.2},
 and per particle (b)
for  different values of $\nu$. The thick solid line in (b) corresponds to the
ideal homogeneous Bose gas given by \re{A.9}. The boundary between BEC and normal phases
is at $t=1$.
\label{fig6ab}}
\end{figure}

    The $\lambda $ - shape behavior of the specific heat per particle for non-interacting bosons for $\nu=1$ has been predicted in Ref. \ci{das}. Below the case with also large $\nu$ will be considered.
 The specific heat per site $C_v/N_s$ and per particle $C_v/N$ calculated from Eqs.
 \re{5.2} are presented in Fig.s \ref{fig6ab} (a) and (b) respectively.
 It is seen that $C_v(\nu,t)/N_s$ reaches unity at rather small values of
  $t$, say,  $t\leq 0.2$ for $\nu\geq 5$ in the BEC phase. In the  normal phase the same quantity
 fast decreases with increasing the temperature that again confirms a weak dependence
 of the energy on $t$, especially at high temperatures. Moreover at
 such temperatures the dependence of the  function $C_v(\nu,t)/N_s$ on $\nu$
 nearly vanishes (see Fig.~\ref{fig6ab}(a)) and it mostly becomes a function
 of only $t$, similarly to the specific heat of IHB gas.
 Therefore $\lambda$ - shaped $C_v/N$, i.e. the specific heat per particle
, decreases as $C_v(\nu,t)/N\sim 1/\nu$ with the  increasing  $\nu$ and fades out
 at large $t$, namely
at $t\geq 2.3$ in contrast to IHB gas, as it is seen from Fig.\ref{fig6ab}(b).

 It is well known that the most of thermodynamic quantities are continuous
 at the critical temperature for IHB gas.
  This remains true for optical lattices also, as it is seen
 from figures Figs.3-Figs.6 .  However there is a discontinuity in
  the slope of  the specific heat, $\partial C_v/\partial T$, even for
 the case of ideal homogeneous gas given by equation \re{1.2}.
 % (see Appendix A).

%%%%%%%%%%%%% FIGURE 6A,B %%%%%%%%%%%%%%%%%%%%%%%%%%
%%%%%%%%%%%%%%%%%%%%%%%%%%%%%%%%%%%%%%%%%%%%%%%%%%%%%%
%%%%%%%%%%%%%%%%%%%%%%%%%%%%%%%%%%%%%%%%%%%%%%%%%%%%%%%

\begin{figure}[H]
\begin{minipage}[H]{0.49\linewidth}
\center{\includegraphics[width=1.1\linewidth]{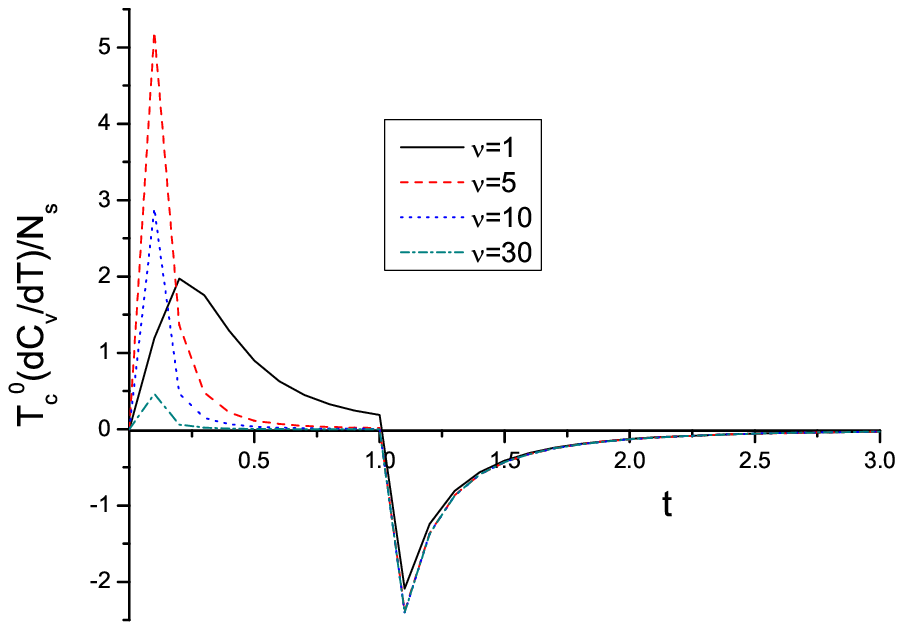} \\ a)}
\end{minipage}
\hfill
\begin{minipage}[H]{0.49\linewidth}
\center{\includegraphics[width=1.1\linewidth]{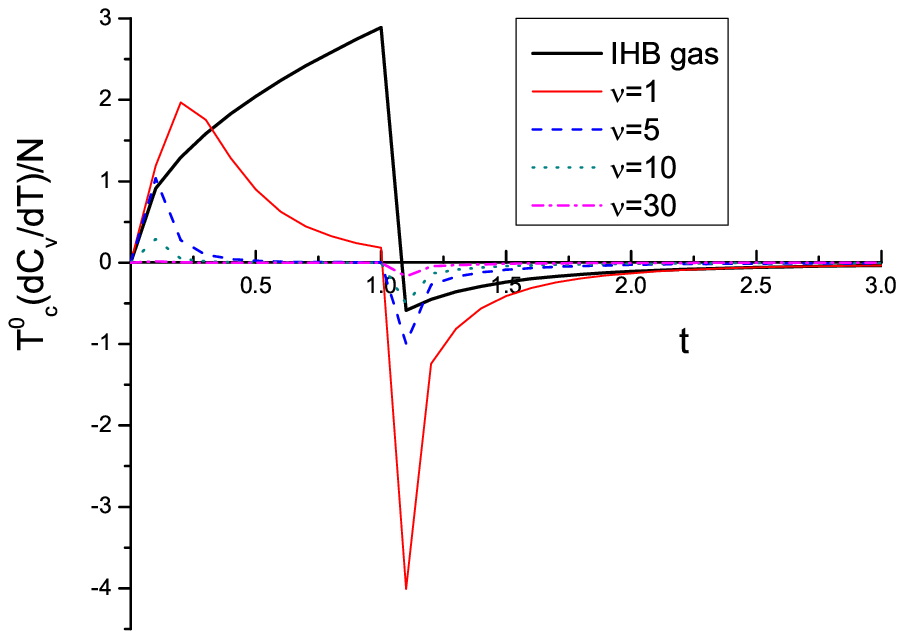} \\ b)}
\end{minipage}
\caption{(Color online)
The temperature derivative of the specific heat per site (a) and
per particle (b)
multiplied by factor $\Tc0$  vs reduced temperature t
for  different values of $\nu$. It is seen that this quantity
for the optical lattices does not reach its maximum at $t=1$,
even at $\nu=1$ (thin solid lines),
in contrast to the ideal homogeneous gas, presented as
a thick solid line in (b)  (see Eq.s \re{A.dcv}).
 }
\label{fig7ab}.
\end{figure}

   The similar quantities, namely, $\partial C_v/\partial T$ per site and per particle
   for noninteracting Bose gases in optical lattices
   are presented in figures Fig.\ref{fig7ab}(a) and Fig.\ref{fig7ab}(b) respectively. It is seen that the derivative is discontinuous in this case also.
  Below  we will show how this jump can be evaluated. First we note that the jump in IHB gas
  presented in Eqs. \re{1.2} and \re{A11} as
  \be
  {\tilde \Delta }   =  \dsfrac{9g_{3/2}(1)  }{4g_{1/2}(1) }+
  \dsfrac{27g_{3/2}^{2}(1)g_{-1/2}(1)  }{8g_{1/2}^{3}(1) }
  \lab{8.2}
  \ee
  is mainly determined by the singularities of Bose functions near $z=1$ which may be isolated by using Robinson formula \ci{robinson}
  \be
  g_\sigma (z)=\dsfrac{\Gamma (1-\sigma)}{\alpha^{2(1-\sigma)}}+regular \quad terms
  \lab{8.3}
  \ee
  where $z=\exp (-\alpha^2)$, $\sigma<1$ , as it was outlined in Appendix A. So, the first term in \re{8.2}
  vanishes, while the second term gives a finite value $\tilde \Delta=3.66$.

  For ideal optical lattices from equations \re{5.3} and \re{5.4} one obtains
  \be
  \ba
  \nu{\Delta} (\nu)= \dsfrac{{\Tc0}}{N_s}\left [ \left ( \dsfrac{\partial { C}_v}{\partial T}\right )_{   \Tc0^{-}} -
  \left (\dsfrac{\partial { C}_v}{\partial T}    \right)_{  \Tc0^{+}}\right]=
             \dsfrac{R_{112}(1)[6R_{223}(1)-2R_{112}(1)-3R_{212}(1)]}{R_{012}(1)}+
         \\
   \\
         \dsfrac{2R_{112}^{2}(1)[R_{112}(1)-3R_{123}(1)]}{R_{012}^{2}(1)}
        +2R_{112}^{3}(1)
        \left
        [
    \dsfrac{R_{023}(1)}{R_{012}^{3}(1)}
  \right
  ]
  \lab{8.4}
  \ea
  \ee
As it was shown in the Appendix B, $R_{ijk}(z)|_{z\rightarrow 1}\rightarrow\infty$
   due to the infrared divergency for $i-k+1<0.$ Thus in Eq. \re{8.4},
    where the relation in square brackets is evaluated in the Appendix B,  only the last term
   survives.
   So, using Eq. \re{b.7} we  get following expression
   \be
   \nu\Delta (\nu)=32\pi^2\left[\dsfrac{R_{112}(1) J}{\Tc0}
   \right]^3.
   \lab{8.5}
   \ee

   At  first glance it seems  that this function behaves like $1/\nu^3$
   due to the Eq. \re{6.2} . However, taking into account the $\nu$
   dependence of $R_{112}(1)$ it can be shown that $\nu\Delta(\nu)\sim const$. In fact, especially,
   for large $\nu$, $(\nu\geq5)$
using the estimation for $R_{112}(1)$ given in \re{R112f} one may conclude that
$\nu\Delta(\nu)$ does not practically depend on the filling factor, since in this case
$\Tc0$ in eq. \re{8.5} is cancelled:
\be
\nu\Delta (\nu)\approx 32\pi^2\left[\dsfrac{\Tc0 J}{4\Tc0 J}\right]^3 =4.9348
   \lab{nudelf}
   \ee
    Actually, performing  exact  numerical calculations
     \footnote{These estimations have been made  at the  points very close to the critical temperature, namely, at  $t_{-1}=0.9890$ and $t_{+1}=1.010.$}
     using \re{8.4} show  following values:
      $\nu\Delta (\nu)\mid_{\nu=1}=3.96$,
$\nu\Delta (\nu)\mid_{\nu=5}=4.618$,
$\nu\Delta (\nu)\mid_{\nu=10}=4.657$ and
 $\nu\Delta (\nu)\mid_{\nu=30}=4.661$  (see also Fig.~\ref{fig7ab}(a)).
 Hence one may conclude that the jump in the heat capacity per particle linearly  decreases with increasing
 the filling factor i.e $\Delta \sim 1/\nu$ as it can be also seen from Fig.~\ref{fig7ab}(b).
  This is in contrast to the case of IHB gases where the similar quantity does not depend on the density being a constant.

 There is one more difference between the $\partial C_v/\partial T$ of  these two kinds of gases.
 As it is seen from  Figs.~\ref{fig7ab} it reaches its maximum exactly at the critical temperature for
 IHB gases, while in the case of optical lattices the maximum is shifted towards smaller values of temperature, $T_{max}<\Tc0$. Moreover, as it is seen from
  Fig.7a, $T_{c}^{0}(\partial C_v/\partial T)/N_s$ is  insensitive to $\nu$
  in the normal phase.
    These facts may be checked experimentally \ci{ruddell} by
 decreasing the interaction between the atoms with a Feshbach resonance technique.

 %\edc
 %%%%%%%%%%%%% FIGURE 7A,B %%%%%%%%%%%%%%%%%%%%%%%%%%
%%%%%%%%%%%%%%%%%%%%%%%%%%%%%%%%%%%%%%%%%%%%%%%%%%%%%%
%%%%%%%%%%%%%%%%%%%%%%%%%%%%%%%%%%%%%%%%%%%%%%%%%%%%%%%

\section{Condensate fluctuations in the thermodynamic limit}

Number-of-particle fluctuations
define
the stability of the system and its way of reaching the
state of thermodynamic equilibrium \ci{fluc1,fluc2}.
 For IHB gases they have been thoroughly
studied by Yukalov in Refs. \ci{yukpre,yuklas}. Here we outline the main
ideas of these works and then discuss
the case of noninteracting gases in cubic optical lattices.

The number-of-particle fluctuations are characterized
by the dispersion
\be
\Delta_{f}^{2}=\langle{\hat N^2}\rangle-\langle{\hat N}\rangle^2
\lab{9.1}
\ee
where ${\hat N}$ is the number - of - particle operator.
This dispersion is directly related to the isothermal compressibility
defined by
\be
\kappa_T\equiv \dsfrac{1}{B}=-\dsfrac{1}{V}\left(\dsfrac{\partial V}{\partial P}\right)_{TN}=
\dsfrac{1}{\rho^2}\left(\dsfrac{\partial \rho}{\partial \mu}\right)_{TN},
\lab{9.2}
\ee
where $B$ is a bulk module, by following equality
\ci{yukkniga}
\be
\kappa_T=\dsfrac{\Delta_{f}^{2}}{TN\rho}
\lab{9.3}
\ee
 A necessary
condition for a system to be stable is the semi-positiveness
and finiteness of the compressibility, that is, $0\leq \kappa_T<\infty$.
If the compressibility \re{9.2} were infinite, this would mean
that an infinitesimal fluctuation of pressure $P$ would lead
to an immediate collapse or explosion of the system.
Therefore, in the thermodynamic limit, the dispersion \re{9.1}
should behave as
\be
\Delta_{f}^{2}\simeq const \cdot N \quad \quad (N\rightarrow \infty)
\lab{9.4}
\ee
When the stability condition \re{9.4} is satisfied, the number of
particle fluctuations are called normal, but when Eq.
\re{9.4} is not valid, so that the compressibility \re{9.3} diverges
in the thermodynamic limit, the fluctuations are termed anomalous.

Now following work \ci{yukpre} it is easy to show that the ideal homogeneous Bose
gas, below the condensation temperature, as an unstable system with anomalous condensate
fluctuations. In fact, as it was shown in the Appendix A,
the compressibility of IHB gas is given by
\be
\tilde\kappa_T=\dsfrac{g_{1/2}(\tilz)}{\lambda_{T}^{3}T\tlrho^2}\quad.
\lab{9.5}
\ee
 In the BEC phase the fugacity
equals to unity, $\tilz(T\leq \tilde \Tc0)=1$, and hence, due to the divergence of
Bose function $g_{1/2}(1)$ the compressibility goes to infinity , $\tilde\kappa_T\rightarrow \infty$.

On the other hand it was shown many years ago by Politzer \ci{politzer} that a magnetic trap stabilizes the system of noninteracting bosons
whose dispersion became proportional to the number of particles:
\be
\Delta_{f}^{2} (mag. trap)=1.37N\left(\dsfrac{T}{\Tc0}\right)^3
\lab{9.6}
\ee
The natural question arises, if  the optical trap with the periodic potential \re{1.1}
is also able to make an ideal Bose gas stable?

 Actually,
in this case the compressibility may be defined similarly to \re{9.2}
as follows
\be
\kappa_T=\dsfrac{a^3}{\nu^2}\left(\dsfrac{\partial \nu}
{\partial \mu}\right)_{TN}
\lab{9.7}
\ee
%  \item {\bf Compressibility} of optical lattice may be directly calculated by using formula
%        given in  \ci{Yukalovobsor}:
 %       \be
  %      \kappa_T=\dsfrac{a^d}{\nu^2}(\dsfrac{\partial \nu}{\partial \mu})
        %\ee
        Representing \re{4.1} as
        \be
        \nu=\int_{0}^{1}dq_1 dq_2 dq_3 \dsfrac{1}{e^{\beta \epsq}z^{-1}-1}
        \lab{9.8}
        \ee
        and differentiating this equation with respect to $\mu$
        we obtain
        \be
        B^{-1}=\kappa_{T}=\dsfrac{a^3 R_{012}(z)}{zT\nu^2}
        \lab{9.9}
        \ee
        where $B$ is the bulk module of the noninteracting gas in an   optical lattice.
        From Eqs. \re{9.9} and \re{9.3}  it is immediately understood that in
          the BEC regime  the ideal optical
        lattice  has an infinitely large particle fluctuations, i.e. $(\Delta_{f}^{2}/N) (BEC)\rightarrow\infty$ and hence becomes
        unstable, since $\ds\lim_{\mu\rightarrow 0}R_{012}(z(\mu))=\infty $ (see Appendix B).
        This is in good agreement with experimental observations.  For instance,
         Roati et al.         \ci{roatiprl} have shown that the  lifetime of the BEC in the optical trap, which is typically
around 3 s, is significantly  shortened when $a_s$ is extremely decreased.

        One more conclusion concerning the scale properties of these two kinds of gases
        can be made by introducing dimensionless compressibility as $\kappa'_T= \Tc0\kappa_T \rho$.
        For IHB gas this quantity may be represented as $\tilde{\kappa}'=\sqrt{t}g_{1/2}(\tilz)/g_{3/2}(1)$, as it was shown
         in the Appendix A. As to the ideal gas in optical lattice the similar quantity
        is $\kappa'_T=R_{012}(z)/zt\nu$. Again we conclude that the dimensionless compressibility
        does not explicitly depend on the density for IHB gases but it does for optical lattices.

 %
 %%%%%%%%%%%%% FIGURE 8 %%%%%%%%%%%%%%%%%%%%%%%%%%
\begin{figure}[H]
\leavevmode
\center{\includegraphics[width=0.7\linewidth]{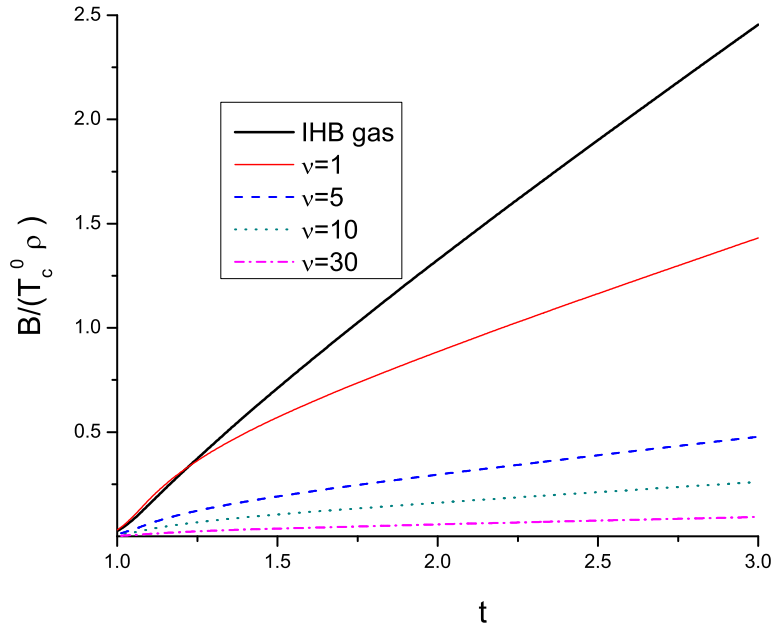} }
\caption{(Color online)
The bulk module in units $\Tc0\rho$   vs reduced temperature t
for  different values of $\nu$. In the BEC phase, when $t\leq 1$ it equals to zero.
The thick solid line corresponds to IHB gas (see  Eq. \re{A13}).
 }
\label{fig8bulk}.
\end{figure}
%%%%%%%%%%%%%%%%%%%%%%%%%%%%%%%%%%%%%%%%%%%%%%%%%%%%%%
%%%%%%%%%%%%%%%%%%%%%%%%%%%%%%%%%%%%%%%%%%%%%%%%%%%%%%%

        In Fig.~\ref{fig8bulk}  the dimensionless bulk module, as an inverse of $\kappa'_T$, is presented
for IHB gas as well as for ideal optical lattice. It is seen that in the normal phase
the module , and hence the compressibility is positive and finite. Thus we may complete this section  with following conclusion : Both kinds of gases under the consideration are
unstable in the BEC phase but stable in the normal phase
\footnote
{
Clearly an ultracold gas, especially in the condensed phase  is  metastable
by itself. Its stability strongly decreases when the interatomic interaction
is switched off.
}.

\section{conclusion}
We have studied thermodynamic properties of ideal gases  loaded into the cubic
   periodic lattice potential in $d=3$ without a harmonic trap for arbitrary integer
   filling factors
   and compared
   them with that of ideal homogeneous Bose gases.
   Although in real experiments a harmonic confinement
   is always present, we hope that our results obtained for    the homogenious case may represent
    a suitable starting point for addressing the trapped problem within a Thomas-Fermi approximation.

   It have been shown by exact numerical calculations that in contrast to
   the case of an IHB gas, the energy as well as the   entropy of ideal optical lattice exhibits a linear dependence
   on temperature in the BEC phase and becomes almost a constant in the normal phase for large filling factors.
   We have evaluated the jump in $dC_v/dT$ and shown that jump in the heat capacity per particle linearly  decreases with increasing
 the filling factor i.e. $\Delta \sim 1/\nu$. It is interesting to note that $dC_v/dT$ for an ideal optical lattice
    reaches its maximum
   not at the critical temperature, as it does for IHB gas, but at rather smaller temperature.

    We have shown that
   scaling properties of these two kinds of gases are different. For example
   the thermodynamic potential of IHB gas
   in units of the critical temperature may be presented as an explicit
   function of only the reduced temperature as $\Omega/\Tc0\sim \Omega'(t)$,
   while that of the ideal optical lattice may be not. Moreover,
   the well known relation $E=-d\Omega/2$ between the energy and the thermodynamic potential
   does not hold for the ideal gas in the periodic potential.

   Studying the bulk properties of an ideal gas in the thermodynamic limit have shown that both kind of gases
   are  unstable
   below the critical temperature . So it seems  impossible to create a stable ideal gas of atoms even with Feshbach resonance technique. For this reason any calculation concerning an ideal gas should be considered as a model one. Nevertheless, such calculations are useful. For example they have already been used to estimate the shift of 
    critical temperature $\Delta T_c=(T_c-T_{c}^{0}) $ of real gases due to the interaction and finite size effects
     \ci{blakiepra76}, or a disorder potential \ci{lopatin}. For real gases at high filling region 
     $\Delta T_c$ may be estimated by using $1/\nu$ expansion \ci{fischer} or effective potential
     methods \ci{teichmann,axel}

    The present work will give an opportunity for
   the estimations in future  experiments  and QMC calculations for large values of  $\nu$ and may serve as
   a check point in theoretical studies in the limit $(U/J) \rightarrow 0 $.

    %The study ideal atomic Bose gases in the normal phase
    %is on progress.
\section*{Acknowledgments}
                    This work is supported by Scientific and Technological Research Council of Turkey (T\"{U}BITAK),  under Grant B\.{I}DEB -2221.

%\newpage

\section*{Appendix A }
\def\theequation{A.\arabic{equation}}
\setcounter{equation}{0}
Here we bring a summary of main explicit formulas for ideal homogeneous  Bose gases .
The most of them can be directly derived from eq. \re{3.1} by using well known thermodynamic
relations \ci{landau,huang}.
\begin{itemize}
        \item {\bf The density}:
        \be
        \rho=\dsfrac{N}{V}=
        \sum_k \dsfrac{1}{z^{-1}e^{\beta \veps(k)}-1       }=\dsfrac{g_{3/2}(z)}{\lambda_T^3}
        \lab{A.1}
        \ee
        where $\sum_k f(k)=V\dsint d {\bf k} f(k)/(2\pi)^3$,  $\veps(k)={\bf k}^2/2m$,
          $\lambda_T=\sqrt{2\pi/Tm}$ and \ci{robinson}
        \be
        \ba
        g_{\sigma}(z)=\dsfrac{1}{\Gamma(\sigma)}\int_{0}^{\infty} \dsfrac{x^{(\sigma-1)}
  dx}{\exp (x)z^{-1}-1}=\Gamma (1-\sigma) {\tilde \alpha}^{\sigma-1}+
          \dssum_{l=0}^{\infty} \dsfrac{(-1)^l}{l!}\zeta (\sigma-l){\tilde \alpha}^l\quad ,\\
          {\tilde \alpha}=-\beta \mu=-\ln z
        \lab{A.2}
        \ea
        \ee
        is the Bose function satisfying the recurrency formula
        \be
        \dsfrac{\partial g_\sigma (z)}{\partial z}=\dsfrac{g_{\sigma-1}(z)}{z}
        \lab{A.3}
        \ee
        \item{\bf The condensed fraction:}
        \be
         n_{0}=1-t^{3/2}
        \lab{n0gaz}
        \ee
        where $t=T/{ \tilde T}_c^{0}$ is the reduced temperature.
        \item {\bf The critical temperature} is defined from eq. \re{A.1} at $z=1$ as:
        \be
        {\tilde  T_{c}^{0}}=\dsfrac{1.05\pi}{m}\rho^{2/3}.
        \lab{A.4}
        \ee
        Here and below  we use numerical values of Bose functions such as $g_{3/2}(1)=2.6124$,
         $g_{5/2}(1)=1.3415$ etc.
         \item {\bf The thermodynamic potential and pressure}. Integration by parts in equation
         \re{3.1} gives
         \be
         \dsfrac{\Omega}{N {\tilde T_{c}^{0}}}=-\dsfrac{PV}{N {\tilde  T_{c}^{0}}}=-\dsfrac{   Vm^{3/2}\sqrt{2} T^{5/2}g_{5/2}(z)      }
         {4\pi^{3/2}{N {\tilde T_{c}^{0}}}
         }\\
         \\
         =\left \{
     \begin{array}{ll}
     -0.513  t^{5/2} & (T\leq {\tilde T_{c}^{0}}) \\
     -0.383  t^{5/2} g_{5/2}(z) & (T> {\tilde T_{c}^{0}})
     \end{array}
   \right.
   \lab{omegagaz}
    \ee

         \item {\bf The fugacity}. Clearly, in the BEC phase $z=\exp(\mu/T)$ equals to unity i.e. $z(T\leq \Tc0)=1$. In the normal phase it can be evaluated as a solution to the
         equation \re{A.1} with a given density $\rho$ and temperature.
           Near the critical temperature,
 when $\mu<<T$ we may use Robinsons formula Eq. \re{A.2} and solve
  \re{A.1} analytically to find following approximation in the normal phase
  \be
  {\tilde z}\approx \exp(-0.54+1.1/t^{3/2}-0.54/t^3) \leq 1 \quad if  \quad  t> 1
  \lab{10.3}
  \ee

%

% vse taki I used linear ALAYAROBINSONlin.mws pologaya .Pri etom alpha^2 rastet s rostom t

          Its temperature derivative $z'=\partial z/\partial T$ can be found by differentiation
          the both sides of eq. \re{A.1} with the fixed $\rho$ with respect to $T$ , i.e. the equation          $d \rho/d T=0$. As a result one obtains
          \be
        %  \ba
          z'{\tilde T_{c}^{0}}=\left \{
     \begin{array}{ll}
          0 & (T\leq {\tilde T_{c}^{0}}) \\
          -\dsfrac{3zg_{3/2}(z)}{2tg_{1/2}(z)} & (T> {\tilde T_{c}^{0}})
          \end{array}
   \right.
                    \lab{A.6}
          \ee
          \item {\bf Entropy and energy}.
          \be
          \dsfrac{S}{N }=\dsfrac{  5 m^{3/2}\sqrt{2} T^{3/2}g_{5/2}(z)      }
         {8\pi^{3/2}\rho
         }\\
         \\
         =\left \{
     \begin{array}{ll}
         1.283  t^{3/2} & (T\leq {\tilde T_{c}^{0}}) \\
      0.957  t^{3/2} g_{5/2}(z) & (T> {\tilde T_{c}^{0}})
          \end{array}
   \right.
          \lab{A.7}
        %  \ea
          \ee
          The energy per particle, when the zero temperature energy is subtracted is given by
          \be
         % \ba
          \dsfrac{E}{N }=
          \left \{
     \begin{array}{ll}
         0.77  { \tilde T_{c}^{0}} t^{5/2} & (T \leq {\tilde T_{c}^{0}}) \\
          0.574  { \tilde T_{c}^{0}} t^{5/2} g_{5/2}(z) & (T> {\tilde T_{c}^{0}})
          \end{array}
   \right.
          \lab{A.8}
          %\ea
          \ee
          \item {\bf The heat capacity and its slope}.
          The exact expression for the heat capacity per particle $C_v/N$, given
          in textbooks,
          \be
          \ba
          \dsfrac{C_v}{N }
         =\left \{
     \begin{array}{ll}
         \dsfrac{15 \zeta(5/2)}{4\zeta(3/2)} t^{3/2}  & (T \leq {\tilde T_{c}^{0}}) \\
         \\
         \dsfrac{15g_{5/2}(z)}{4g_{3/2}(z)}-\dsfrac{9g_{3/2}(z)}{4g_{1/2}(z)}&(     T> {\tilde T_{c}^{0}})
         \end{array}
   \right.
          \lab{A.9}
          \ea
          \ee
   where $\zeta (x)$ is the Riemann function, may be repalced by a nice and more practical approximation given in ref. \ci{wang}
   \be
          \ba
          \dsfrac{C_v}{N }
         =\left \{
     \begin{array}{ll}
         1.926 t^{3/2}  & (T \leq {\tilde T_{c}^{0}}) \\
         \\
          1.496+0.341t^{-3/2}+0.089 t^{-3}  & (T> {\tilde T_{c}^{0}})
          \end{array}
   \right.
          \lab{A.10}
          \ea
          \ee
   Differentiating the last equation one obtains
   \be
   \ba
          \dsfrac{{\tilde T_{c}}}{N }\left(
          \dsfrac{\partial C_v}{\partial T}
          \right )_V
         =\left \{
     \begin{array}{ll}
         2.899 t^{1/2}  & (T \leq {\tilde T_{c}^{0}}) \\
         \\
          -0.511t^{-5/2}-0.267 t^{-4}  & (T> {\tilde T_{c}^{0}})
          \end{array}
   \right.
          \lab{A.dcv}
          \ea
          \ee
   The discontinuity in the slope of the heat capacity may be evaluated directly from
   the equation \re{A.dcv} as
   \be
   \ba
   {\tilde\Delta}= \dsfrac{{\tilT _c}}{N}\left [ \left ( \dsfrac{\partial {\tilde C}_v}{\partial T}\right )_{   \tilde{T}_{c}^{-}} -
  \left (\dsfrac{\partial {\tilde C}_v}{\partial T}    \right)_{  \tilde{T}_{c}^{+}}\right]
  =
  \dsfrac{9g_{3/2}(1)  }{4g_{1/2}(1) }+
  \dsfrac{27g_{3/2}^{2}(1)g_{-1/2}(1)  }{8g_{1/2}^{3}(1) }=\\
  \\
  \left [
  2.889t^{1/2}+0.511t^{-5/2}+0.267t^{-4}
  \right ]_{t\rightarrow 1}=0.36675
     \lab{A11}
   \ea
   \ee
   \item {\bf The compressibility.}
   Being defined as
   \be
   \kappa_T=\dsfrac{1}{\rho^2}
   \left( \dsfrac{\partial \rho}
    {\partial \mu}
   \right)_{N,T}
   \lab{A12}
   \ee
    the compressibility $\kappa_T$ can be directly obtained from \re{A.1}
    .
    The result is
    \be
    \ba
    \kappa_T \rho \tilde{T}_{c}^{0}=
    \left \{
     \begin{array}{ll}
    \infty  & (T\leq {\tilde T_{c}^{0}})\\
    \\
    0.3828{\sqrt{t}g_{1/2}(z) }   &  ( T > {\tilde T_{c}^{0}})
    \end{array}
   \right.
    \lab{A13}
    \ea
    \ee
    Note that the bulk module is defined as $B=1/\kappa_T$.

                            \end{itemize}

\section*{Appendix  B}
\def\theequation{B.\arabic{equation}}
\setcounter{equation}{0}

Here we consider the functions $R_{ijk}(z)$ defined as

\be
          \ba
          R_{ijk}(z)= \int_{0}^{1}dq_1 dq_2 dq_3
          \dsfrac{
          x^i e^{jx}
          }
          {
          \left [ e^{x}z^{-1}-1 \right ]^k
          },
          \\
          x=\beta \epsq
          \lab{b.1}
          \ea
          \ee
          For IHB gases the similar functions are Bose functions whose expansion
          was given by Robinson \ci{robinson} as in the Eq. \re{A.2}.
          %\be
          %g_\sigma({\tilde \alpha})=\Gamma (1-\sigma) {\tilde \alpha}^{\sigma-1}+
          %\dssum_{l=0}^{\infty} \dsfrac{(-1)^l}{l!}\zeta (\sigma-l){\tilde \alpha}^l
          %\lab{b.2}
          %\ee
   %       where ${\tilde \alpha}=-\beta \mu$, and $\mu$ - the chemical potential.
   From \re{A.2} it is seen that in the BEC regime for ideal gas, when $\mu\rightarrow 0$
   and $\sigma<1$,    $g_\sigma({\tilde \alpha})$ diverges, e.g.
   $g_{1/2}({\tilde \alpha})\sim \sqrt{\pi}/\sqrt{{\tilde \alpha}} $.

   Similarly it is easy to understand that when the chemical potential goes to zero
   the function in \re{b.1} goes to infinity, i.e. $R_{ijk}(z)|_{z\rightarrow 1}\rightarrow\infty$
   due to the infrared divergency for $i-k+1<0$ and is regular otherwise. Below we show how this divergency can be isolated and presented analytically e.g. for $R_{012}(z)$ and $R_{023}(z)$. To do this
   we use Debye like  approximation \ci{ourknr1, Yukalovobsor} :
   \be
   \ba
   \dsint_{0}^{1} dq_1dq_2dq_3 \rightarrow \dsfrac{\pi}{2}\dsint _{0}^{q_d}q^2 dq;
   \\
   \veps (q)=2J\veps'(q), \quad \quad \veps'(q)=\dsfrac{\pi^2 q^2}{2},  \quad \quad
   q_d=\left(\dsfrac{6}{\pi}\right)^{1/3}.\\
\lab{b.3}
\ea
\ee
Introducing $T'=T/J$, $z=\exp(\mu/T)\equiv \exp(-\alpha^2)$ and expanding the exponent,
 we represent
$R_{ijk}(z)$ as
\be
R_{ijk}(z)\mid _{\alpha\rightarrow 0}=\dsfrac{\pi}{2}\dsint _{0}^{q_d}q^2 dq
\dsfrac{x^i e^{jx}}{(x+\alpha^2)^k}
\lab{b.4}
\ee
where $x=\veps(q)\beta=\pi^2 q^2 /T'$.
In particular
\be
\ba
R_{012}(z)\mid _{z \rightarrow 1}=\dsfrac{(T')^{3/2}}{2\pi^2}
\dsint _{0}^{y_d}y^2 dy
\dsfrac{1+y^2}{(y^2+\alpha^2)^2}\\
\\
R_{023}(z)\mid _{z \rightarrow 1}=\dsfrac{(T')^{3/2}}{2\pi^2}
\dsint _{0}^{y_d}y^2 dy
\dsfrac{1+2y^2}{(y^2+\alpha^2)^3}\\
\\
R_{001}(z)\mid _{z \rightarrow 1}=\dsfrac{(T')^{3/2}}{2\pi^2}
\dsint _{0}^{y_d} dy
\dsfrac{y^2}{(y^2+\alpha^2)}\\
\lab{b.4}
\ea
\ee
where $y_d=\pi q_d/\sqrt{T'}$. The explicit integration gives
\be
\ba
R_{012}(z)\mid _{z \rightarrow 1}=\dsfrac{(T')^{3/2}}{4\pi^2}
\left [
\dsfrac{y_d(2y_{d}^{2}+3\alpha^2-1  )} {y_d^2+\alpha^2}
-\dsfrac{(3\alpha^2-1)\arctan(y_d/\alpha)}{\alpha}
\right ]\\
\\
R_{023}(z)\mid _{z \rightarrow 1}=\dsfrac{(T')^{3/2}}{16\pi^2}
\left [
\dsfrac{y_d(- \alpha^2-6\alpha^4+y_{d}^{2}-10y_{d}^{2}\alpha^2  )} {\alpha^2(y_d^2+\alpha^2)^2}
+\dsfrac{(6\alpha^2+1)\arctan(y_d/\alpha)}{\alpha^3}
\right ]\\
\\
R_{001}(z)\mid _{z \rightarrow 1}=\dsfrac{(T')^{3/2}}{2\pi^2}
\left [
y_d-{\alpha\arctan(y_d/\alpha)}
\right ]\\
\lab{b.5}
\ea
\ee

Now expanding in powers of $\alpha$ one obtains

\be
\ba
R_{012}(z)=\dsfrac{(T')^{3/2}}{8\pi\alpha}+ regular \quad terms\\
\\
R_{023}(z)=\dsfrac{(T')^{3/2}}{32\pi\alpha^3}+\dsfrac{3(T')^{3/2}}{16\pi\alpha}+ regular \quad terms\\
\\
R_{001}(z)=\dsfrac{T'6^{1/3}}{2\pi^{4/3}}-\dsfrac{(T')^{3/2}\alpha}{4\pi}+
\dsfrac{(T')^{2}6^{2/3}\alpha^2}{12\pi^{8/3}}+O(\alpha^3)
\lab{b.6}
\ea
\ee
%%%%%%%%%%%  see  Alya Robinson .mws
where the regular terms are finite at $\alpha\rightarrow 0$ and may be calculated
more accurately  by the exact  three dimensional integration.
As to $R_{001}(z)$ which will be used  below  to study the fugacity
$z=\exp(-\mid\mu\mid/T)$ near $\Tc0$,
it is regular at small $\mid\mu \mid/T\equiv \alpha^2$, as expected.

Now we are on the stage of calculating the relation
$\ds\lim_{z\rightarrow 1}  R_{023}(z)/R_{012}^{3}(z) $ which is necessary to evaluate the
jump in the heat capacity. From equation \re{b.6} one immediately obtains
\be
  \lim_{z\rightarrow 1}  \dsfrac{R_{023}(z)}{R_{012}^{3}(z)}=16\pi^2\left[\dsfrac{ J}{\Tc0}\right]^3
  \lab{b.7}
  \ee
  .

For completeness we estimate also $R_{112}(1)$ which is used to evaluate the discontinuity
in the slope of the specific heat in eq. \re{8.4}. In fact, for large $\nu$, $(\nu\geq 5)$
one may represent  $R_{112}(1)$ as
\footnote{Although this  simple approximation is not valid for
 the system of  homogenous
 atomic gases, it is justified for optical lattices due to the fact that
 $\tilde{\varepsilon}(q)$ is bounded above i.e. $\mid \tilde{\varepsilon}(q) \mid \leq 2d$.  }
\be
\ba
R_{112}(1)= \dsint_{0}^{1}dq_1 dq_2 dq_3
          \dsfrac{
          \veps(q)\beta e^{\veps(q)\beta}
          }
          {
          \left [ e^{\veps(q)\beta}z^{-1}-1 \right ]^2
          }\approx \Tc0 \dsint_{0}^{1}\dsfrac{dq_1 dq_2 dq_3}{\veps(q)}=
          \dsfrac{\Tc0}{2J} \dsint_{0}^{1}\dsfrac{dq_1 dq_2 dq_3}{\tilde\veps(q)}
\lab{R112}
\ea
\ee
where $\tilde\veps(q)= \dssum _{\alpha=1}^{3}(1-\cos\pi q_\alpha)$
Now  evaluating the last integral numerically gives following final expression :
\be
\lab{R112f}
R_{112}(1)\approx
          \dsfrac{\Tc0}{4J}
\ee

 Similarly to IHB gas one may obtain an approximation for the fugacity
 of an ideal  optical lattice, starting from following  equation
  \be
  \nu=\dsint_{0}^{1}
  \dsfrac{dq_1dq_2dq_3}{z^{-1}e^{\beta \epsq}-1       }\equiv R_{001}(z)
  \lab{10.3}
  \ee
  Near the critical temperature, $\alpha$, defined through $\alpha^2=-\ln(z)=\mid \mu\mid/T$,
  is small, so one may use Robinson like expansion \re{b.6} for $R_{001}(z)$
  to solve \re{10.3} analytically. As a result we obtain
  \be
\ba
\alpha\approx\dsfrac
{
2(6/\pi)^{1/3}
}
{
(C_\nu t \nu)^{1/2}
}
-
\dsfrac
{
4\pi\nu
}
{
(C_\nu t \nu)^{3/2}
}\quad ,\quad z=\exp(-\alpha^2)
\ea
\ee
where $C_\nu=3.96\exp(0.37/\nu)$ and $T$ is presented as $T=t\Tc0$, with $\Tc0 $ given by \re{6.2}.
From this equations as well as from Fig.\re{fig4abcd}(c) one may conclude that
the fugacity goes to unity with the increasing of the filling factor $\nu$.
As to $z'=dz/dT$ defined in equation \re{4.5} it is clear that $z'(T)\mid _{T=T_{c}^{0+}}=0$.
Thus, near $\Tc0$ both $z$ and $z'$ are continuous.

 \newpage
 \bb{99}
 \bi{feshbach} E. Timmermans, P. Tomassini, M. Hussein and A. Kerman,
 Phys. Rep. 315 (1999) 199.
 \bi{roatiprl} G.  Roati et al. Phys. Rev. Lett. 99(2007) 010403.
 \bi{Roatinature} G. Roati et al. Nature 453(2007)895.
 \bi{8roati} Y. Shin et al., Phys. Rev. Lett. 92(2004)050405.
 \bi{9roati} L. Pezz{\'{e}} et al., Phys. Rev. A72(2005) 043612.
 \bi{bailier80} D. Baillie and  P. B. Blakie, Phys. Rev. A80(2009)
031603(R) .
 \bi{bailie80} D. Baillie and P. B. Blakie, Phys. Rev. A80(2009) 033620.
 \bi{lewenstein} M. Lewenstein, A. Sanpera, and V. Ahufinger, {\it
Ultracold atoms in optical lattices: Simulating quantum many-body
systems} ,Oxford University Press, Oxford, 2012.
\bi{NCOOPER} N. R. Cooper, Phys. Rev. Lett.,106 (2011) 175301 .
\bi{zwerger}W. Zwerger, J. Opt. B5 (2003)S9
\bi{stoofbook} H.T.C. Stoof, K.B. Gubbels, and D.B.M. Dickerscheid,
{\it Ultracold Quantum Fields} ,Springer, Berlin, 2009.
\bi{Yukalovobsor} V. I. Yukalov, Laser Physics  19(2009)1 ;\\
V. I. Yukalov, Condensed Matter Physics, 16(2013) 23002.
 \bi{landau} L. D.  Landau and E. M.  Lifshitz, {
\it E. M. Statistical Physics}, Part  1  Oxford, England:
Pergamon Press, 1980
\bi{huang}
Kerson Huang,
{\it Introduction to Statistical Physics} Second Edition, CRC Press, 2001.
\bi{wang} Frank Y.-H. Wang, Am. J. Phys.  72(2004)1193
\bi{mancarella} F. Mancarella, G. Mussardo and
 A. Trombettoni, Nucl. Phys. B887 (2014) 216  .
 \bi{ourannals} A. Rakhimov, S. Mardonov, E.Ya. Sherman, Ann. Phys. 326(2011) 2499 .
   \bi{matsumoto}M. Matsumoto,  B. Normand,  T. M. Rice, and M. Sigrist,
   Phys. Rev. B69(2004) 054423 .
   \bi{weli}  W. Li, A.K.  Tuchman, H.C. Chien and M.A. Kasevich, Phys. Rev. Lett. 98  (2007) 040402.
   \bi{polkovnikov} I. Danshita, A. Polkovnikov, Phys. Rev. A 84 (2011) 063637.
   \bi{hajibaba} Z. Hadzibabic , S. Stock , B.  Battelier ,V.  Bretin , J.  Dalibard,
   Phys. Rev. Lett. 93  (2004) 180403.
   \bi{bloch} S. Foelling, A. Widera, T. Mueller, F. Gerbier, I. Bloch
Phys. Rev. Lett. 97 (2006) 060403. 
\bi{ourknr2} H.Kleinert, Z. Narzikulov, A. Rakhimov,
Journal of Statistical Mechanics: Theory and Experiment Vol. 2014,(2014) P01003 .
\bi{ourknr1} H. Kleinert, Z. Narzikulov, A. Rakhimov Phys. Rev. A85(2012)
063602 .
\bi{pathria}R. K. Pathria, {\it Statistical Mechanics}, 2nd ed. Butterworth-Heinemann,
Boston, 1996.
\bi{ouriman1} A. Rakhimov and I. N. Askerzade, Phys. Rev. E90(2014) 032124.
\bi{keterle} W. Ketterle and N. J. van Druten, Phys. Rev. A54(1996)656.
\bi{pra69} P. B. Blakie and J. V. Porto, Phys. Rev. A69(2004)013603.
\bi{das} R. Ramakumar and A. N. Das, Phys. Lett. A348 (2006)304.
\bi{robinson} J. E. Robinson Phys. Rev. 83 (1951) 678.
\bi{ruddell} S. K. Ruddell, D. H. White, A. Ullah, M. D. Hoogerland, arXiv:1409.5494 (2014)
\bi{fluc1} D.J. Evans and D.J. Searles, Adv. Phys. 51(2002) 1529 .
\bi{fluc2}  R. Dewar, J. Phys. A36, (2003)631 .
\bi{yukpre} V. I. Yukalov Phys. Rev. E72(2005)066119 .
\bi{yuklas} V. I. Yukalov Laser Phys. Lett. 1 (2004) 435 .
\bi{yukkniga}V. I. Yukalov, {\it Statistical Green's Functions},
Queen's University, Kingston, 1998.
\bi{politzer} H. D. Politzer, Phys. Rev. A54 (1996)5048.
\bi{blakiepra76} P. B. Blakie1, and Wen-Xin Wang, Phys. Rev.  A76 (2007)  053620. 
\bi{lopatin} A.V. Lopatin and V. M. Vinokur, Phys. Rev. Lett. 88 (2002) 235503. 
\bi{fischer} Uwe R. Fischer, Ralf Schützhold, and Michael Uhlmann, Phys. Rev. A 77, (2008) 043615 . 
\bi{teichmann} N. Teichmann, D. Hinrichs, M. Holthaus, A. Eckardt, Phys. Rev. B 79 (2009) 100503.
\bi{axel} F. E. A. dos Santos and A. Pelster, Phys. Rev. A 79 (2009) 013614.
\eb
 \edc